\documentclass[sn-apa]{sn-jnl}

\usepackage{graphicx}%
\usepackage{multirow}%
\usepackage{amsmath,amssymb,amsfonts}%
\usepackage{amsthm}%
\usepackage{mathrsfs}%
\usepackage[title]{appendix}%
\usepackage{xcolor}%
\usepackage{textcomp}%
\usepackage{manyfoot}%
\usepackage{booktabs}%
\usepackage{algorithm}%
\usepackage{algorithmicx}%
\usepackage{algpseudocode}%
\usepackage{listings}%

\usepackage{geometry} 
\newcommand{\repitem}{%
  \addtocounter{enumi}{-1}%
  \let\savedtheenumi\theenumi
  \renewcommand{\theenumi}{\arabic{enumi}\itemasterisk}%
  \item\let\theenumi\savedtheenumi
}
\protected\def\itemasterisk{\rlap{*}}

\newcounter{saveenumerate}
\makeatletter
\newcommand{\enumeratext}[1]{%
\setcounter{saveenumerate}{\value{enum\romannumeral\the\@enumdepth}}
\end{enumerate}
#1
\begin{enumerate}
\setcounter{enum\romannumeral\the\@enumdepth}{\value{saveenumerate}}%
}
\makeatother
\usepackage{xurl}

\theoremstyle{thmstyleone}%
\theoremstyle{thmstyletwo}%

\theoremstyle{thmstylethree}%

\begin{document}

\title[Article Title]{The Change-Driver Account of Scientific Discovery: Philosophical and Historical Dimensions of the Discovery of the Expanding Universe}


\author[1,2]{\fnm{Patrick M.} \sur{Duerr}}\email{patrick-duerr@gmx.de}

\author[3]{\fnm{Abigail} \sur{Holmes Mills}}\email{amills63@gatech.edu}

\affil[1]{\orgdiv{Faculty of Philosophy}, \orgname{University of Oxford}, \orgaddress{\street{Radcliffe Humanities, Woodstock Road}, \city{Oxford}, \postcode{OX2 6GG}, \country{UK}}}

\affil[2]{\orgdiv{Martin Buber Society of Fellows for Advanced Research in the Humanities and Social Sciences}, \orgaddress{\street{Hebrew University, Mt Scopus Campus}, \city{Jerusalem}, \postcode{91905}, \country{IL}}}

\affil[3]{\orgdiv{School of Public Policy}, \orgname{Georgia Instiute of Technology}, \orgaddress{\city{Atlanta}, \state{Georgia}, \country{USA}}}


\abstract{
What constitutes a scientific discovery? What role do discoveries play in science, its dynamics and social practices? The paper explores these questions by first critically examining extant philosophical explications of scientific discovery---the models of scientific discovery, propounded by Kuhn, McArthur, Hudson, and Schindler. As an alternative, we proffer the ``change-driver model”. In a nutshell, it conceives of discoveries as problems or solutions to problems that have epistemically advanced science; here we take a problem to be generated by a datum that we want to account for and make sense of---by putting it in contact with our wider web of scientific knowledge and understanding. The model overcomes the shortcomings of its precursors, whilst preserving their insights. We demonstrate its intensional and extensional superiority, especially with respect to the link between scientific discoveries and the dynamics of science, as well as with its reward system. Both as an illustration, and as an application to a recent scientific and political controversy, we apply the considered models of discovery to one of the most momentous discoveries of science: the expansion of the universe. We oppose the 2018 proposal of the International Astronomical Union as too simplistic vis-à-vis the historical complexity of the episode, and as problematically reticent about the underlying---and in fact crucial---philosophical-conceptual presuppositions regarding the notion of a discovery. The change-driver model yields a more nuanced and circumspect verdict: (i) The redshift-distance relation shouldn’t be named the ``Hubble-Lemaître Law”, but “Slipher-Hubble-Humason Law”; (ii) Its interpretation in terms of an expanding universe, however, Lemaître ought to be given credit for; (iii) The Big Bang Model, establishing the expansion of the universe as an evidentially fully warranted result in the 1950s or 1960s (and a communal achievement, rather than an individually attributable one), doesn't qualify as a discovery itself, but was inaugurated by, and in turn itself led to, several discoveries.
}




\maketitle

\section{Introduction}\label{sec1}

Discoveries indubitably lie at the heart of science. Staple examples comprise a diverse array: the phases of Venus, gravitational waves, Archaeopteryx, the Lascaux Cave Paintings, the double helix structure of the DNA, $E=mc^2$, complex numbers, isomerism, X-rays, continental drift, HIV as the cause of AIDS, Lorentz attractors in chaos theory, Wiles’ proof of Fermat’s Last Theorem, or the Immortal Jellyfish’s immortality. \par

Intertwined with the notion of a scientific discovery are major descriptive and historiographical, as well as normative and epistemological-methodological questions (see e.g.\ Arabatzis, 2006, pp.220). Discovery figures centrally in key facets of science: its internal dynamics, with discoveries viewed as the shining achievements of science (think of the Human Genome Project, or the images of Black Holes) and hallmarks of progress (think of penicillin, or mind-boggling accuracy of discoveries in particle physics), and several of its social practices, with honours (such as the Nobel Prize) heaped upon illustrious discoverers, and pedagogy, with scarcely any textbook or popular science story not revelling in the adventures and vicissitudes of science’s heroes, discoverers. \par

But what exactly does ``discovery" mean? What does the notion include? What does it exclude? Should we regard the term as a passe-partout concept, applicable to anything that might preoccupy a person’s intellectual/mental life, e.g.\ artistic inspiration, stumbling on a piquant photo of the Queen’s manicurist, a solution to a crossword puzzle, a philosophical or theological brainchild, any scientific or technological break-through or what have you? Or ought we, at least in the context of science, to reserve the term for a more specific usage? If so, how to construe it? And how is it related to the various features of science?\footnote{Nickles (2000, p.85) provides—without answering them—a list of salient questions: ``(w)hat constitutes a scientific discovery? How do we tell when a discovery has been made by whom and whom to credit? Is making a discovery (always) the same as solving a problem? [...] May discovery require a long, intricate social process? Could it be an experimental demonstration? How do we tell exactly what has been discovered, given that old discoveries are often recharacterized in very different ways by succeeding generations? What kinds of items can be discovered, and how? Is the discovery of a theory accomplished in much the same way as the discovery of a new comet, or is discovery an inhomogeneous domain of items or activities calling for quite diverse accounts? Must a discovery be both new and true? How is discovery related to (other?) forms of innovation, such as invention and social construction?”}\par
Given the ubiquity\footnote{Of late, the need for conceptual clarification has become especially acute. The notion of scientific discovery—or rather: the idea of an individual discoverer within the commonsense way of thinking about discoveries—has come under pressure in at least two regards. First, we live in an era of Big Data science; research is often conducted within large-scale collaborations.  Credit, such as in the form of awards, is often not given to these collaborations as a whole, but rather to a select few individuals. The Nobel Prize committee, in particular, limits its prizes to three people. Can large collaborations as a whole be discoverers? Can present accounts of discovery make sense of the large scale and collaborative nature of much of modern science? Secondly, the use of AI in science, leading to discoveries, raises the question of agency: whom to ascribe agency in the discovery to? Is the AI just a tool (with whom as its user?) or may it qualify as a discoverer sui generis? 
While these intriguing questions lie outside of the present paper’s ambit, it's clear that an explication of the notion of a scientific discovery will be an essential prerequisite for satisfactory answers.} of talk about discoveries in science, one is dumbfounded by the paucity of philosophical attention devoted to it (cf. Michel, 2022, especially sect.1\&2): attempts at conceptually clarifying the notion are bafflingly rare. One might perhaps hope to find a more fleshed-out reconstruction of it and the role of discoveries in science in the loci classici---Popper’s (1983; 2002a; 2002b) \emph{Logic of Scientific Discovery}, Hanson’s (1958) \emph{Patterns of Discovery}, and Kuhn’s (1996) \emph{The Structure of Scientific Revolutions}. Alas, one would look there somewhat in vain.\footnote{The otherwise excellent \emph{Stanford Encyclopedia} entry (Schickore, 2022) is another case in point. As is Nickles (2000), despite raising the most pertinent questions (see fn.1)! An exception is Blackwell’s (1969) largely neglected monograph, on which we’ll draw for our account in \textbf{§3}.} Relying on strong assumptions and being unduly restricted in their domain of applicability, the few recent examples (inspected in \textbf{§2}) turn out to be unpersuasive. 

The present paper seeks to develop a more satisfactory, philosophical model of the notion in its multifarious dimensions, hinted at above: our goal is to provide an explication of discovery---in the sense of Carnap (1950, pp.5): a simple, intensionally and extensionally adequate, and fertile precisification.\par

The discovery of the expanding universe marks the birth of modern cosmology. The techniques developed to measure the expansion remain essential to cosmology. They drove the later, and only slightly less revolutionary discovery of our own times: that of ``dark energy"—phenomena associated with the accelerated expansion of the universe (see e.g. Durrer, 2011). In spite of its profound importance---or quite possibly because of it---the attribution of credit for the insight that the universe is undergoing expansion continues to be a topic of ongoing debate. Historical accounts of the discovery (e.g. Kragh \& Smith, 2003 or Nussbaumer \& Bieri, 2009) primarily impugn the received wisdom: they dispute the previously recognised discoverers, Edwin Hubble and Alexander Friedmann, and forefront the contributions of Georges Lemaître. Likewise, in 2018 the International Astronomical Union voted to rename the former ``Hubble Law” (which characterises the present expansion rate of the Universe) to the ``Hubble-Lemaître Law”. This was done to honour Lemaître's role in the discovery. Yet, without a more rigorous philosophical conception of discovery, it’s difficult to argue where the credit is due—or why getting the record straight even matters. The significance of this discovery, and the continuing debate over how to distribute credit, make the expanding universe a fruitful case for philosophical analysis.\par

We will proceed as follows. \textbf{§2} will critically examine the extant models of discoveries, found in the literature, beginning with Kuhn’s account and continuing with philosophical elaborations to it. Having distilled from those analyses lessons about what a philosophical explication of the notion should achieve—and what to shun—we’ll present and discuss, in \textbf{§3}, an alternative, the ``change-driver model''. \textbf{§4} will apply it to the historical case of the discovery of the expanding universe—and juxtapose our model’s verdict with those of the other accounts. The main findings of our paper are summarised in \textbf{§5}.

\section{Extant models of discovery}\label{sec2}
This section will take stock of the main accounts of scientific discovery in the literature. We'll critically examine in turn Kuhn’s (\textbf{§2.1}), Hudson’s (\textbf{§2.2}), McArthur’s (\textbf{§2.3}), and Schindler’s (\textbf{§2.4}) models. \textbf{§2.5} will cull general shortcomings of all existing accounts. This will cast into sharper relief the desiderata of a more satisfactory model, sketched in \textbf{§3}.  

\subsection{Kuhn: discoveries and the dynamics of science}\label{Sec2Kuhn}

Kuhn’s oeuvre (1962; 1977; 1996) is one of the earliest systematic explorations of discovery. Subsequent analyses, including those that we’ll inspect in \textbf{§2.2-§2.4}, build upon his ideas. It’s therefore apposite to commence our review of accounts of discovery with Kuhn.\par

Kuhn distinguishes between two types of scientific discoveries: ``what-that" and ``that-what" discoveries (in Schindler’s (2015) evocative terminology). In \textbf{``what-that"-discoveries}, a theoretical framework (which, notoriously, Kuhn refers to as a paradigm) predicts—or induces the expectation of—a new fact or phenomenon: \emph{what} is supposed to exist precedes the empirical confirmation \emph{that} it exists. Classic cases of predictions fall into this class: ``the neutrino, radio waves, and the elements which filled empty spaces in the periodic table” (Kuhn, 1962, p.761). By contrast, in \textbf{``that-what”-discoveries}, such as that of Uranus (observed by a number of astronomers, on multiple occasions, but eventually classified/identified as a planet by Lexell, see Kuhn, 1996, pp.115) or that of X-rays, the temporal order is reversed: a fact or phenomenon is observed in advance of its satisfactory conceptualisation within a theory or framework; thus the discovery isn’t expected.\par

Each type impacts science in different ways. ``What-that”-discoveries occur with a theoretical-conceptual framework in place (in Kuhn’s historiographical model: within ``normal science”\footnote{``(N)ormal science, an enterprise that, as we have already seen, aims to refine, extend, and articulate a paradigm that is already in existence” (Kuhn, 1996, p.122). In the parlance of Kuhn’s postscript (op.cit., pp.174), successful ``what-that”-discoveries figure centrally in the elaboration of the paradigm’s disciplinary matrix.}). They yield ``mere additions or increments to the growing stockpile of scientific knowledge” (Kuhn, 1962, p.763). By corroborating anticipations—expectations buttressed by other successes of the framework—they solidify and deepen knowledge. Usually, the assimilation of ``what-that”-discoveries by the adopted theoretical framework requires the latter’s further refinement, e.g. the development of new computational techniques, more detailed or de-idealised models, etc. Successful ``what-that”-discoveries thus afford a ``measure of progress” (ibid.), achieved within the theoretical framework.\par

Much more viscerally ``that-what”-discoveries ``act back upon what has previously been known […]” (ibid.). They have the potential for greater disruption or ``subversion” (op.cit, p.5; 1962, p.362): ``(a) discovery like that of oxygen or X-rays does not simply add one more item to the population of the scientist's world” (Kuhn, 1996, p.7). A ``sense that something was amiss” (op.cit., p.56) dawns on the scientist—a mismatch between the received theoretical framework and the phenomenon. ``That awareness of anomaly opens a period in which conceptual categories are adjusted until the initially anomalous has become the anticipated” (op.cit., p.64). The anomaly’s ``assimilation” or its ``digestion” (Lakatos) by the framework—the phenomenon’s conceptualisation—requires profound conceptual and theoretical re-adjustments in the existing theories. Such discoveries can thereby trigger scientific revolutions, a “qualitative transformation and quantitative enrichment of fundamental novelties” (cf. ibid.).\par

Kuhn’s treatment of discoveries has a surprising, and indeed untoward, consequence: as far as ``that-what”-discoveries (Schindler, 2015, pp.125) are concerned\footnote{``(O)nly when all the relevant conceptual categories are prepared in advance, in which case the phenomenon would not be of a new sort, can discovering that and discovering what occur effortlessly, together, and in an instant” (Kuhn, 1996, pp.55). In other words, ``what-that”-discoveries often lack a significant time dimension; they do admit of uniquely identifying discoverers.
}, ``[…] any attempt to date the discovery or to attribute it to an individual must inevitably be arbitrary. Furthermore, it must be arbitrary just because discovering a new sort of phenomenon is necessarily a complex process which involves recognizing both that something is and what it is” (1962, p.762; verbatim: 1996, p.55). Not only is the recognition of what a phenomenon is a temporally extended process. According to Kuhn, it also implicates a blanket refashioning of the conceptual-theoretical background. Such a process tends to be convoluted, with contributions from sundry agents (cf.\ Arabatzis, 1996).
In this regard, Kuhn makes three claims:  
\begin{enumerate}
\item[(1)] the absorption of discoveries is marked by a murky and prolonged conceptual-theoretical re-adjustment process; 
\item[(2)] the ``distinction between discovery and invention, or between fact and theory” (1996, p.52) is “exceedingly artificial”—in other words: empirical discovery is theory-laden, with an inseparable mixing of theoretical and empirical elements;
\item[(2*)] the (semantic) incommensurability of paradigms undermines the possibility of comparing scientific claims couched in different frameworks, due to their radical shifts in meaning.
\end{enumerate}
(1) usually complicates the unambiguous spatio-temporal localisation of a discovery act (and a discoverer). It’s a practical hindrance for the working historian—not an in-principle impediment. Likewise (2) is merely a practical challenge: an enlightening comparison of theoretical descriptions that utilise different conceptual frameworks is typically a non-trivial task. It often requires judicious analysis;\footnote{For a successful example of such a comparison---in fact a rebuttal of one of Kuhn’s explicit examples for (2*)---see Beisbart \& Jung (2004).} occasionally even new conceptual-semantic or mathematical-formal tools need to be developed.\footnote{For an example---again a rebuttal of one of Kuhn’s examples for (2*)---see Fletcher (2019).} On both points, (1) and (2), we largely concur.  With the more fine-grained taxonomy of discoveries, presented in \textbf{§3.2}, we’ll show that these tenets of Kuhn’s—which, given their fairly uncontroversial nature, we’ll treat as philosophical insights—can be retained, without inescapably curtailing the ability to ascertain when and by whom a “that-what”-discovery was made. \par
For Kuhn, the clincher for this latter claim pivots on (2*). Conceptualisations, he opines, are prerequisites for a discovery. As per (2), they are inseparably entrenched in a theoretical framework. Rivalling frameworks, however, differ so radically that those conceptualisations lack a common measure; semantically, they are mutually opaque. With their conceptualisations being incommensurable, talk of discoveries (both ``that-what"-discoveries, as well as ``what-that"-discoveries) that belong to, or inaugurate, different theoretical frameworks/paradigms is inane; a discovery is inherently framework-indexed (meaningfully defined only relative to a framework). In order to coherently talk about variegated discoveries in one breath, one must adopt the same theoretical framework for their conceptualisation. As a historian, though, Kuhn is loath to single out any such framework; to privilege those of our present state-of-the art, in particular, smacks of anachronism and historiographical blunder. Consequently, a discovery  remains eo ipso a perspectival construct. In short, Kuhn’s incommensurability thesis strips the notion of a scientific discovery of an objective sense: for the historian—as a matter of principle—it’s impossible to unequivocally study discoveries. This is surely a rebarbative conclusion (indeed one that Kuhn doesn’t spell out explicitly). It hinges on (2*), Kuhn’s incommensurability thesis. The authors discussed in \textbf{§2.2}-\textbf{§2.4} forgo it, and rescue an objective notion of discovery, by jettisoning the incommensurability thesis. Instead, they postulate a form of continuity (of varying strengths) between successful theories. \par

Blackwell (1969, p.86) rightly “(emphasises) that Kuhn does not intend to develop a theory of discovery as such. His is the wider question of the growth and development of science”. Herein, we maintain, lies the most important insight of Kuhn’s analysis: the connection between discoveries and the progressive dynamics of science. Discoveries and the modus operandi of science, its historical evolution, are essentially interlaced. We regard this as a key lesson---one that other models of discovery \emph{don’t} sufficiently heed. The notion of a scientific discovery must, we urge, be embedded in the dynamics of science.\par
This is precisely what Kuhn accomplishes with his distinction between the two types of discoveries (as Schindler (2015, p.124) rightly observes). Each plays a slightly different role within the dynamics of science (see especially Kuhn, 1962, pp.363). Decisively for this link, Kuhn ties the notion of a discovery to some kind of conceptualisation: “observation and conceptualization, fact and assimilation of fact, are inseparably linked, in the discovery of scientific novelty.” (1962, p.762). “What-that”-discoveries presuppose a prerequisite framework; the discovery must, more or less, fit into its conceptual-theoretical resources. Any residual mismatch is the source of innovation: via modifications, additions, and/or refinements within the existing framework, one seeks to minimise that mismatch. Thereby, the successful resolution makes the framework more sophisticated, enhancing its accuracy, range of applications, versatility, etc. “That-what”-discoveries, by contrast, burst the confines of the old framework: relative to the latter, they are anomalous. The resolution must come from a new framework: if the tension between the existing framework and the discovery becomes too pronounced, it “(necessitates) paradigm change” (op.cit., p.58). The “that-what”-discovery has shaken, and precipitated a re-configuration of, the web of orthodox belief; it has thereby advanced science in a fundamentally more innovative way than its “what-that”-counterpart. \par

Later commentators on Kuhn (\textbf{§2.2}-\textbf{§2.4}) tie the notion of a discovery to a more demanding kind of prerequisite conceptualisation—problematically so (as we’ll see). Our alternative model of discovery (\textbf{§3}) shows how Kuhn’s insight into the intimate connection between discoveries and the dynamics of science can be implemented in a different way (without rendering the dating of a certain kind of discovery, and the identity of its discoverer, \emph{necessarily} futile).

\subsection{Hudson: Materially demonstrated base descriptions}\label{sec2Hudson}
On Hudson’s (2001) model of scientific discovery, four criteria must be satisfied for the discovery of an object X (by an individual, the discoverer D): 

\begin{enumerate}
    \item \textbf{Base description}: D must provide what Hudson calls a “base description” for X. This denotes “a description of the object that suffices to identify it: something that satisfies this description is the object being considered” (p.77). 
    \item \textbf{Material demonstration}: D “succeeds in demonstrating materially that this base description is satisfied” (p.79).
    \item \textbf{Novelty condition}: D “finds something which is novel relative to a particular social community [...]” (ibid.).
    \item \textbf{Truth condition}: “The object described and materially demonstrated is, \emph{in fact}, X [...]” (ibid., our emphasis)
\end{enumerate}

A few comments are in order. First, the level of detail and accuracy required of a base description “is not hard and fast” (p.78); it varies from (historical) context to context. What matters, Hudson argues, is that given the knowledge available to D, a base description can fulfil two functions. First, it serves as a means of (re-)identification, as an indicator for X: “(t)o have discovered an object one need \emph{only possess enough conceptual resources to recognize its presence in a fairly reliable manner}, and such resources are what base descriptions provide for us” (p.78, our emphases). The same dependence on D’s context, the background knowledge of their time in particular, applies to the level of accuracy required of the material demonstration. A second function of the base description is to instil in D at least an idea of her discovery’s significance: “(Hudson) (takes) the act of discovery to be a reflective event, one that impresses itself upon the discoverer as significant and informative” (ibid.).\par

Secondly, Hudson (p.81) stresses “[...] the contextualized nature of the novelty condition: novelty is socially dependent [...]. Thus, priority will sometimes depend on who gets included in the relevant ambient social group.” For Hudson, the novelty condition is a temporal absolute, detached from social realities: it doesn’t require that D’s work and priority actually be recognised. \par

Thirdly, how do commonplace discoveries (of, say, the watch of one’s grandfather in the attic) differ from scientific discoveries? According to Hudson (op.cit., pp.81, our emphases), “(s)uch \emph{commonplace items are familiar} to us and well-understood, whereas \emph{scientific kinds on their initial discovery are quite unfamiliar} and sometimes even exotic in their properties [...]. [...] We are admitting that scientific discoveries are \emph{not of a different kind} than commonplace discoveries, but \emph{differ only in the order of complexity} regarding the sorts of base descriptions that are used and the sorts of material demonstrations that need to be performed” (pp.81).\par

Although Hudson's model has useful features, we ultimately find it to be untenable.\par

Commendably, the model aspires to a natural desideratum: unlike Kuhn’s model, it seeks to clearly label individuals as discoverers. This practice, together with the usually attendant bestowal of honours and rewards, belongs to the social aspects of the notion (as we’ll elaborate in \textbf{§3.3}).\par

However, Hudson’s model asks for both too little and too much. First, Hudson’s flexibility towards what a base description must deliver exacts the price of vagueness. Is there a limit not only to how accurate in detail a discoverer’s base description must be, but also to how much inaccuracy (or how much falsehood in the description) is still tolerable (cf.\ Schindler, 2015, p.128)? Analogous worries can be voiced about the material demonstration. Hudson’s minimal answer—that the base description and material demonstration should suffice to rule out a sufficient number of relevant alternatives (p.78)—is hand-waving. It flies in the face of both contemporary and historical queries (e.g. Laudan, 1981; Psillos, 1999, Ch.5 \& 6; cf. McArthur, 2011, sect. 3). Amongst the latter, the luminiferous ether stands out as theoretically indispensable prior to 1905, and likewise amply “materially demonstrated” (see e.g. Torretti, 2009). Hudson doesn’t take up this historical challenge.\par


At the same time, Hudson’s model asks for too much: why insist that a discoverer D possess \emph{both} a sufficiently accurate base description and material demonstration? Sometimes, as in the case of gravitational waves, discoveries can be—prima facie—purely theoretical, with their material demonstrations (i.e. empirical confirmations) significantly delayed. Vice versa, as in the case of ultra-high energy rays in the cosmic radiation, sometimes empirical discoveries precede an adequate “base description”.\footnote{Bartels (2021, Ch.3.7) discusses the illuminating example of a phenomenological discovery, still awaiting its accurate “base description”: Alzheimer’s disease.} It strikes us as plausible to treat both kinds as \emph{distinct} discoveries in their own right (as \textbf{§3.2} will elaborate).\par

We are also unhappy with Hudson's limitation of his model to the discovery of objects. This contravenes bona fide examples of discoveries concerning other types of entities: superconductivity (a \emph{property} of certain materials), isospin (a quantum number related to a \emph{symmetry} of nucleons), or seafloor spreading (a geological \emph{process}, respectively). One may also be wary of the restriction to \emph{material} objects (cf.\ also McArthur, 2011, pp.368): that implausibly precludes otherwise bona fide discoveries from the mathematical and theoretical sciences, e.g.\ the discovery of the Casimir effect of quantum field theory (op.cit., sect.4).\par

Thirdly, Hudson expressly allows for the continuity between commonplace discoveries and scientific ones. A modicum of continuity seems indeed an attractive feature of the model. Yet, his demarcation criterion strikes us as wrong-footed: familiarity—apart from being yet another problematically woolly notion—is too contingent a \emph{psychological-subjective} condition to base the distinction between scientific and non-scientific discoveries on.\footnote{While perhaps the demarcation between science and non-science can’t be drawn \emph{rigidly} (cf.\ Laudan, 1996, Ch. 11), we don’t believe that it lacks any basis other than a psychological-subjective hunch, such as “familiarity”.} \par

Finally, Hudson rightly underlines the relevance of a discovery’s social context. Via the novelty condition, Hudson’s model indeed achieves an important desideratum: it rules out that, say, undergraduates in chemistry lab classes discover oxygen. Regrettably, his incorporation of a discovery’s social dimension remains unsatisfactory. What makes one community more suitable, or relevant, than another: one’s academic school/tradition? one’s tribe? one’s country? one’s language community? What about “independent researchers”? Hudson intimates awareness of the problem. As a remedy, he proposes that ``the ambient social group is the entirety of the human race, or more generally, all Earth-bound rational beings over the course of all time” (p.81, our emphasis). To our minds, however, it detracts neither from the scientific value or significance nor the impressiveness of a discoverer’s achievements, if in another part of the world her discovery had been made by someone else—so long as their two communities and cultures are sufficiently isolated. Hudson’s novelty condition appears to be an unduly strong fiat. In the same vein, Hudson’s model precludes, by definition, simultaneous discoveries—a pervasive phenomenon in science.\footnote{ It even has a wikipedia-entry of its own: \url{https://en.wikipedia.org/wiki/List_of_multiple_discoveries}!} Again, this consequence seems unduly prohibitive. It rubs against common practice (where simultaneous, independent discoverers are typically given \emph{shared} credit). \par

\subsection{McArthur: Discoveries for the structural realist}\label{sec2McArthur}
McArthur (2011) “(works) out some revisions to [Hudson’s] account by drawing from a structural realist view of theory change” (p.361). 
Structural realists (of the epistemic stripe, see e.g. Votsis, 2020) contend that scientific knowledge primarily resides in the patterns and relationships amongst observable entities, information encoded in a theory’s structural-mathematical content, its equations. Viewing its structural content as what is “essential” (McArthur, 2011, p.371) to a base description, McArthur relaxes Hudson’s demands on the discoverer’s epistemic grasp of it: rather than a “\emph{literally}” (van Fraassen, 1980, esp. Ch. 1\&2) approximately true base description, she possess a base description that is structurally approximately true. \par

More in detail, for an individual D to qualify as the discoverer of an entity $X$
(an object, process, event, etc.),  McArthur stipulates three conditions:
\begin{enumerate}
    \item \textbf{Base description}: D must possess a base description of X (in the form of a minimal theoretical description/conceptualisation).
    \item \textbf{Novelty condition}: D must be the first to provide such a base description. 
    \item \textbf{Structural adequacy condition}: This base description for X must be structurally adequate; its relevant structural content has already been corroborated. 
\end{enumerate}

Many features—and shortcomings—carry over from Hudson’s model. We'll zoom in on one. McArthur rightly (given our discussion in \textbf{§2.2}) remarks that Hudson’s model is predicated on ``an inadequate account of theory change" (p.366): it presupposes an interpretation of the base description in scientific-realist terms, assuming that the base description deserves to be understood as (approximately) “\emph{literally} true”. As a structural realist, McArthur bypasses that problem: structural realism posits a weaker sense of conceptual and referential continuity between past and present science. The structural content of empirically and explanatorily successful theories, structural realists avow, survive theory change.  \par
Because of his commitment to structuralism, McArthur’s proposal remains unconvincing, though. It sensitively relies  on a specific (and as such controversial) position within the scientific realism debate. 
Here, we won’t rehearse the general complaints about epistemic structural realism (see e.g. Psillos, 1999, Ch.7, 2001; Chang, 2003; Ainsworth, 2014). In the present context, one problem in particular stands out: structural realism is limited to theories admitting of a \emph{sufficiently mathematised} form. But mathematised theories outside of the mathematical-physical sciences are rare! Should we therefore, as McArthur’s model would seem to imply, really deny the possibility of many discoveries in, say, biology, psychology or medicine? This seems preposterous.\par

\subsection{Schindler: Discovery as essence identification}
Schindler's (2015) model strengthens Hudson’s account of discovery by demanding that the base description capture, at least in part, the discovered phenomenon’s essence:
\begin{quote}
A discovery of X requires \textbf{observing X or its direct effects} […] and the \textbf{correct conceptualisation of those of X’s \emph{essential properties} that suffice (epistemically) to individuate X} at time t […] (op.cit., p.132, emphases in the original).
\end{quote}

For a discovery, the phenomenon’s (materially demonstrated) base description—in Hudson’s terminology—must contain at least some true beliefs about X’s characteristic features, defining what X is and discriminating X from other conceivable entities with similar effects. Thanks to such robust partial knowledge about X, according to Schindler, the phenomenon has been sufficiently comprehended—both empirically (qua material demonstration) as well as theoretically (qua the identification of its (partial) essence)—to licence the title of a discovery.\par

Through the insistence on essential properties in the base description, Schindler tries to overcome what he perceives as the main shortcoming of Hudson’s model (which Hudson (2001, p.78) in turn animadverts upon in Kuhn’s)—the double “what-indeterminacy”-problem (p.128): how \emph{accurate} does a base description have to be for a discovery? And how much \emph{in}accuracy or \emph{in}completeness is permissible? The identification of some of X’s essential properties is supposed to address the first; the epistemic background—enabling the distinction between X and alternative entities—is supposed to take care of the second question.\par

Let's concentrate on his employment of essential properties. Schindler strongly invites a reading of his proposal at face value (see, inter alia, the literature he cites, see p.132, fn14): he seems committed to essences, as discussed in the metaphysics literature—that is, properties that define an object’s nature or identity, and that the object can’t lack without ceasing to exist. Regrettably, Schindler elucidates neither his understanding of essences nor the kind of essentialism he has in mind. In light of the sizable literature (see e.g. Roca-Royes, 2011; Bird, 2009; 2018; Ishii \& Atkins, 2020) on essentialism, this omission makes Schindler’s proposal tricky to evaluate. But even at a general level, three arguments militate against an appeal to essences in this strong sense.\par

First, one may simply shy away from such a controversial metaphysical notion. In fact, ``it is not obvious how best to characterize the notion of essential property, nor is it easy to give conclusive arguments for the essentiality of a given property” (Roca-Royes, 2011, p.65). A fortiori, one may be leery of tying one’s understanding of scientific discoveries—a high-level notion permeating scientific practice in all disciplines—to such a specific (and notoriously vexed) idea in metaphysics.\par

Secondly, accounts of essentialism typically stipulate a tight definitional link between essentiality and metaphysical necessity (see e.g. Ishii \& Atkins, 2020, esp. sect.1-3). Necessity here is construed in terms of truth in all possible worlds.\footnote{Also Schindler explicitly endorses a (modal) definition of essences: “(p)resumably, there is a list of properties that are (metaphysically speaking) necessary and sufficient for the individuation of electrons”(p.132).} Whether science has much to say about such modal speculations is questionable (cf.\ Norton, 2021). Accordingly, one may doubt the relevance of essentialism for understanding science in general, and scientific discoveries in particular. 

Thirdly, essentialism sits uneasily with the special sciences, such as geology, biology, or social psychology. Due to their inherent complexity, fuzzy conceptual boundaries, variability and dependence on specific circumstances and interests of the inquirer, talk of essential properties for the objects of those disciplines faces well-rehearsed challenges. That, on Schindler’s account, would compromise the possibility of discoveries in those disciplines. This strikes us as inordinately restrictive—as Schindler’s (2015, sect. 4.1) own discussion of a bona fide discovery in geology illustrates, or more dramatically: the discovery of anthropogenic climate change (see Friedrich, 2022).\footnote{Although clearly envisaging the foregoing strong metaphysical reading in terms of essentialism, Schindler also contemplates a weaker reading (p.132., fn14): “(a)nti-essentialists may plug in their preferred notion of natural kinds here.” On this weaker reading, discoveries require the observation of X (or its direct effects), in tandem with a sufficiently correct classification of X’s natural kind. \par

The world, its objects and processes, adherents of natural kinds opine, admit of natural groupings—a division into said natural kinds (see Dupré, 2000; Bird \& Tobin, 2023; Brzović, 2018 for details); the latter are supposed to carve nature at its joints. The most prominent proposal for such classification or categorisation in terms of natural kinds is indeed essentialism: natural kinds are anchored in (sufficiently similar) essences. Other approaches are metaphysically less fastidious and more pluralistic: they conceive of natural kinds as groupings on the basis of clusters of properties, i.e. family resemblance.\par

Again, Schindler’s omission of further details prevents a deeper analysis of how these alternate accounts apply to his model. At first blush, though, it’s far from clear that they are compatible with his further commitment to a “realist stance” (p.123). First, Schindler’s invocation of essences, or natural kinds, is designed to underwrite an \emph{objective} distinction between essential and accidental properties—properties that are necessary for a discovery’s prerequisite conceptualisation, and those that aren’t. One may wonder, however, whether groupings based on family-resemblance (rather than putative essences) can deliver that: instead, the taxonomic lines of demarcation tend to depend on the context (see e.g.\ Ruphy, 2010) —and the \emph{subjective} interests of the inquirers (as is frequently explicitly acknowledged in the non-essentialist approaches to natural kinds). Secondly, and relatedly, Schindler seems implicitly committed to semantic realism: his insistence on “\emph{correct} conceptualisation” of a phenomenon (reiterated passim)—up to “at least some false and incomplete beliefs about X” (p.128)—is naturally read in realist terms (cf. Psillos, 1999, esp. Ch.1)—as an assertion about the (partial) literal truth of a theory’s unobservable theoretical content.
Prima facie, however, such semantic-realist ambitions—whilst underwritten by essentialism—are at odds with a context-dependent classification and categorisation of the world into natural kinds.}  

\subsection{General shortcomings}
We’ll now step back to compile a handful of general inadequacies of the above three post-Kuhnian models of discovery (i.e.\ McArthur’s, Hudson’s, and Schindler’s). By way of negative examples, they’ll sharpen our sense of what a satisfactory model \emph{should} achieve---paving the way for our task in \textbf{§3}.\par

First, the scope of the extant models is expressly \textbf{restricted to material objects}. This a priori limitation is unduly narrow (see also M\"unster, 2022, pp.68). Not only material \emph{objects}, but also properties, processes, events, etc. can plausibly be discovered. By the same token, one should also allow for discoveries of \emph{immaterial} entities, e.g. physical laws, or mathematical theorems.\par

Secondly, all three post-Kuhnian accounts overtly evince \textbf{bias towards realism}; each presupposes substantial realist premises. \par

An alternative that isn’t hostage to such controversial assumptions would be desirable (especially for historians of science, see Arabatzis, 2001): a satisfactory model of scientific discovery ought not to depend on specific positions in the realism/anti-realism debate. \par

One reason for the pro-realist stance lies in warding off a consequence of Kuhn’s account (of which the other three models are intended as improvements): Kuhn’s incommensurability thesis spoils the objectivity of the notion of a discovery across major theory-change. One is then stuck with the catch-22 of having to choose between either a particular philosophical framework, or of swallowing paradigm-relativism about discoveries. Neither option is attractive. The extant accounts seek to dodge the dilemma by blocking the incommensurability via pro-realist assumptions. As we’ll see in the next section, it’s possible—and plausible—to renounce the former, without subscribing to the latter.\par
The commitment to realism has also a second source of motivation. One may likely want to eschew talk of discoveries today discarded as scientifically obsolete (e.g.  the creation of mercuric oxide via calx). By the same token, the pro-realist approach is likely also undergirded by the surmise that what makes a scientific discovery valuable is its enduring value; past scientific achievements would consequently have to exhibit sufficient continuity with it in order to qualify as discoveries. From a scholarly perspective, however, such a preference for \textbf{triumphalist presentism} (“Whiggism”, Butterfield, 1931) is disconcerting; it’s a dubious historiographic doctrine (and the sustained target of Kuhn’s writings, see Hoyningen-Huene, 2012). It’s desirable that a model of discovery overcome such a bias. The challenge is to achieve this whilst doing justice to the intuitions that drive the two arguments.\par

Thirdly, the three philosophical accounts pay \textbf{insufficient attention to two key facets of scientific discoveries}:
\begin{enumerate}
    \item \textbf{Their connection with theory dynamics}: how are discoveries and the advancement of science related? How do discoveries and science interact?
    \item \textbf{Their communal dimension}, and the various social-institutional practices surrounding it: what makes discoveries scientifically valuable?\footnote{Note the entanglement of this question and our above comments on the Whiggish bias of the extant philosophical accounts.} Why do textbooks extol the achievements of discoverers?  What rationale—if any—underlies the idea of awards and other signs of recognition for discoverers?
\end{enumerate}

Discoveries, for Kuhn, are the pi$\grave{e}$ces de r$\acute{e}$sistance for the evolution of science. The other models' advocates reject his overall historiographical model. Nonetheless, they hold on to Kuhn's account of discoveries---by trying to improve on it. Vis-$\grave{a}$-vis this tension, they remain largely silent on (1). But how do discoveries and the dynamics of science influence each other? Those authors follow Kuhn, however, in setting aside the questions under (2). They explicitly elide the reward system of science in particular. Presumably, such issues are relegated to the sociology of science—and viewed as a haphazard, messy, extraneous element of science, of little concern to epistemology.\par

Neglecting those features of discoveries is, we contend, remiss: an explication of the notion of a discovery ought to reconstruct both a natural link between discoveries and theory dynamics, as well as between discoveries and the communal-sociological dimension of science.\par

Finally, the tension between the goal of emending Kuhn’s model, and the rejection of Kuhn’s larger framework for the dynamics of science spawns another problem (closely related to the preceding point (1)): \textbf{why} do Hudson, McArthur or Schindler \textbf{weld the notion of a discovery to the need for the discovered phenomenon’s conceptualisation?} What speaks against (more or less) \emph{purely} observational discoveries, or (more or less) \emph{purely} theoretical-conceptual-interpretative discoveries? Within Kuhn’s two-phase framework of science, the link is clear: no scientific result without a paradigm, either the received one, or a new and revolutionary one. But having (rightly!) distanced oneself from Kuhn’s framework, why insist on a sufficiently detailed  (and accurate!) conceptualisation as a sine qua non for a discovery?

\section{The change-driver model: discoveries as drivers of scientific advances}\label{SecChangedriver}
This section will canvass an alternative account of scientific discoveries, the “change-driver model”. \textbf{§3.1} will introduce the model. \textbf{§3.2} will present a more fine-grained typology of scientific discoveries which naturally arises within it. We’ll then (\textbf{§3.3}) discuss the merits of the model. Finally, in \textbf{§3.4}, we’ll inspect, and rebut, some challenges.

\subsection{The change-driver model of scientific discovery}\label{sec3sub1}
Combining ideas of Popper (1994; 2002a; 2002b) and Kuhn (1962; 1996), we propose to define discoveries as particular kind of problem-solving/posing (see also Gutting, 1980 and Koertge, 1982). We thereby locate discoveries at the heart of science's dynamics.

With its emphasis on how discoveries propel science, we’ll dub our account the ``change-driver model". It elaborates Blackwell’s (1969) remark that ``(t)he word `discovery' is a rather generic term that includes quite diverse instances of the advancement of human knowledge” (p. 96). For that, “to be counted as a discovery, [a new idea] must be integrated in the accepted body of scientific knowledge. This acceptance may be friendly or disruptive, depending on the impact of the new idea on former views" (p.53). What those ``diverse instances" have in common is, we maintain, that they either solve or pose problems (pace Arabatzis, 2006, p.219).\par

More rigorously, we stipulate that a scientific discovery is
\begin{enumerate}
\item[(1)] a cognitive scientific result R (observational/experimental finding, explanation, theorem, prediction, etc.) that
\item[(2)] has significantly advanced the scientific field in question in the following triple sense:

\begin{itemize}
\item[(P)] R either solves or poses a sufficiently well-defined problem; and: 
\item[(J)] sufficient justification for R has been proffered; and:
\item[(A)] R is what we'll call ``epistemically ampliative".
\end{itemize}

\end{enumerate}

Let's unravel these conditions in more detail.

The seemingly innocuous condition (1) delineates the admissible \emph{objects} of a discovery. In principle, they may include: empirical results (e.g. the detection of the Cosmic Microwave Background), proofs (e.g. Poincaré’s discovery of the chaotic dynamical behaviour in the three-body problem of celestial mechanics), an application of a theory that yields novel predictions (e.g. the application of basic quantum mechanics and general relativity to white dwarfs or neutron stars), the articulation of a new and powerful conceptual or taxonomic framework (e.g. Mendeleev’s periodic table), etc. The condition licences the discovery of material or immaterial entities, objects, structures, properties, states, or processes. No apriori restrictions are imposed on a discovery's ontological category. \par

The rider ``cognitive” requires that the scientific result directly pertain to science’s \emph{cognitive} aims: it constitutes a leap forward in scientific understanding and knowledge. Scientific discoveries are supposed to be intrinsically valuable for science, rather than instrumentally so. By contrast, technological or practical accomplishments (say, the refraction telescope, or a quantum computer) \emph{promote} science's cognitive aims.\footnote{The distinction between cognitive and technological results lies on a continuum (see e.g.\ Douglas, 2014; Shaw, 2022). This parallels the blurry line that marks practical from theoretical knowledge.} Note that this excludes---at least as discoveries for the \emph{empirical} sciences---new calculational or observational techniques (e.g.\ Feynman diagrammes or mass spectrometry). Scientifically valuable as they are, they are better viewed as powerful tools. 

Scientific discoveries, we maintain, represent a particular kind of cognitive achievement; they are special ways in which the cognitive aims of science are realised. What distinguishes them is what the advance clause (2) seeks to capture. 
\par

It prescribes that scientific discoveries impact the dynamics of science: not only do discoveries constitute cognitive achievements; due to their particular \emph{interaction} with it, they also significantly advance science. The three conditions, (P), (J), and (A) spell out the intended specific kind of progressive change that discoveries enable.\par

(P) is concerned with a scientific result's fit---consonance vs. disruption---with a background web of scientific beliefs. A scientific discovery R poses or solves a problem, or ``puzzle". (Although aptly capturing the tightly constrained, and/or constraining, nature of discoveries, we desist from using the latter term because of its Kuhnian overtones.)
Either the existing body of knowledge and understanding naturally makes sense of---absorbs or subsumes---R (with perhaps minor adjustments), or not. In the latter case, R makes one scratch one's head: hard work is necessary to relieve the perplexity that R creates. 

We deem it insightful to phrase this in terms of posing or solving a(n explanatory/explicative) \emph{problem} in a more precise sense (following Nickles, 1978, 1980, 1981). Generated by a (theoretical or observational, see \textbf{§3.2}) datum $X$, a problem is composed of two key ingredients. The first is the problem-question: how to account for $X$? How to comprehend it? How to explain (i.e. place it in a wider web of causal and other dependency relations) or accommodate/conceptualise $X$ (i.e. integrate it into a conceptual-theoretical framework)? The problem-question problematises the datum's fit into our wider web of scientific knowledge and understanding.

The second ingredient of a problem consists in the constraints on answers to its problem-question. They fall into two principal sets (cf. Meheus \& Bartens, 1996).\footnote{We gloss over their inclusion of methodological instructions, ``(specifying) the operations we should or should not fulfil in order to reach the solution, or at least to come closer to it" (p.163); they include also heuristic considerations. We consider these instructions contained in the requirements (J): the condition encodes standards of epistemic warrant and kosher reasoning.} The first is given by the pertinent background knowledge that the scientific community, in the main, accepts: the scientific assumptions (``certainties", op.cit., pp.161) that aren't questioned. Consistency and coherence (of varying degrees) with this background narrow down answers to an account of $X$: they delimit possible solutions of the problem. This ``orthodox" background fulfils two further functions. One is to ``determines the meaning of the entities" (op.cit., p.162) that the datum $X$ and possible answers may invoke. Another is to ``partly determine the operations that are considered as justified" (ibid.), inferential and justificatory practices that, against the communal background knowledge, are considered compelling. 

A second set of constraints form less entrenched hypotheses. These ``relevant statements" (Meheus \& Batens) don't belong to the generally accepted background; albeit more conjectural and tentative, they shouldn't contradict it, either. The problem-solver brings them to bear on the problem in an innovative, often original way. (Of course, it's incumbent on her to substantiate them. If a problem or its solution is to count as a discovery, this is precisely what our condition (J) requires.) Whereas the constraints from background knowledge determine possible solutions to a problem, the ``relevant statements" impose restrictions on the latter in order to whittle down the options to the \emph{right} solution. ``Right" or ``correct" (Meheus \& Batens, 1996, p.162) here isn't supposed to be necessarily cashed out in realist terms. Rather, we take it to be a correlative requirement of the justification condition (J). A discovery must be sufficiently---by reasonable epistemic standards---unambiguous: either its problem question or its answers, subject to the various constraints, ought to be more or less uniquely defined. 

For a result R to satisfy the clause (P), the scientific finding R must either generate or solve a problem in the just-mentioned sense. For the second condition, (J), constraints must be in place that make the problem sufficiently well-posed (minimally, ensuring that an answer is expected to exist\footnote{The clause, expressing a form of minimal coherence with background knowledge, precludes what we have reason to regard as brute fact coincidences (such as those conspiracy theories traffic in, but also less pathological speculations such as Dirac's Large Number Hypothesis).}), with the constraints having garnered sufficient epistemic-evidential warrant. More on this below.



A scientific result R conforming to (P) makes one wonder: either, does R satisfy a sufficiently tight and secure set of constraints from our background knowledge? If so, R is said to solve a problem in our sense. Alternatively, when R generates a problem, one asks: does R add new constraints, or challenges or revises older constraints from that background knowledge---with the hunt now on for a finding that meets the new set of constraints?

In our account of discovery, the condition (P) serves two functions. First, it implements the thought that discoveries dynamically and vigorously interact with a field: they hang together with its body of cognitive achievements in a way that reverberates more widely. Discoveries that prompt a question challenge our background knowledge; discoveries that satisfactorily solve a problem enrich it.

A second motivation for (P) stems from the etymologically\footnote{In many languages, the respective words for ``discovery" have the roots of un-covering something hitherto concealed (see Strobach, 2022).} encapsulated hunch that a discovery discloses something pre-existing, but previously hidden. With (P), the time was ripe for the discovery; the answer to a problem was in the air. Or, the discovery brought an issue to light that had, unnoticed, been lurking in the background. Either way, discoveries, through their characterisations via problems ``are viewed as objective entities. They 'exist' at some historical point, in a knowledge system and related practice. Whether they are or are not discovered by someone is irrelevant to their existence" (Meheus \& Batens, 1996, p.156). Below, we'll see how this intuition underwrites the demarcation of discoveries from more free-wheeling---but by no means necessarily nugatory---scientific activities.   

``Advance" is a success term---as is ``discovery". In both cases, the success in question concerns a result's epistemic status. This the second condition in (2) aims at implementing. (J) requires that a scientific discovery be backed up by appropriate epistemic warrant. It can't just be a \emph{wild guess}, or a piece of science fiction. Good epistemic---typically: evidential---arguments must make a scientific result sufficiently plausible. Combining this with the foregoing condition (P), a discovery presupposes sufficient reason to believe that it (re-)solves (or raises) a genuine problem.   

Note that what counts as sufficient justification for (J) isn't intended to be fixed or rigid/context-independent. To specify the epistemic-evidential standards that one deems appropriate should, to our minds, be left to the discretion of the historian who examines specific discoveries. We find it preferable to allow for a plurality of voices and perspectives (see e.g. Chang, 2008, 2009, 2017, 2021). Historiography is steeped in normative epistemological presuppositions that guide and structure historical narratives (Nanay, 2010; Dimitrakos, 2021). An account of scientific discovery ought not to dictate historians which normative-epistemological canons to subscribe to. It does, however, encourage  transparency: historians should put their normative-epistemological cards on the table.      

Our motivation for (J) is twofold. One is the already-mentioned success-laden connotation of ``discovery". (J) seeks to rule out specious effects and findings, such as statistical flukes, untethered speculations, poorly conducted experiments or calculations, etc. By reasonable standards, discoveries are supposed to count as epistemic successes. Qua \emph{scientific} discovery, a candidate result should, of course, meet the evidential-epistemic criteria, customary in science. (No sane historian would countenance Sheldrake's ``morphic fields" or Tipler's ``Omega Point" as discoveries.) Our account doesn't dictate, though, whether these standards should be fully \emph{historicised}---those of historical agents---or, say, \emph{our} standards (hoped to enjoy the benefit of hindsight). This again is a decision we leave to the discretion of the historian (exhorting, again, only transparency).\footnote{Depending on how austerely a historian construes the requisite standards for justification/evidence, she might not grant Hawking radiation or the holographic principle (AdS/CFT duality) the status of a discovery, but merely that of an (eminently promising and seminal) idea.} 

A second motivation for (J) derives from the desideratum that a scientific discovery should be important: meeting epistemic standards strengthens the ties between a scientific discovery and background knowledge. Not only does this confer confidence in a result, but it also means that one can't easily disentangle it from a larger body of beliefs. 

It's not always trivial to recognise a scientific result as an answer to a problem, or a problem itself, in a way that meets (J). The 2014 announcement of the BICEP2 microwave telescope that certain patterns in the primordial gravitational wave background had been detected (so-called ``B-modes"): initially hailed as ``one of the most dramatic discoveries in decades" (Livio \& Kamionkowski, 2014, p.8) that confirmed an important prediction of cosmic inflation, the observations were soon debunked as resulting from dust in the Milky Way. As Livio \& Kamionkowski underscore (op.cit., p.10), ``the B-modes story demonstrates how progress in science is truly achieved. Rather than through a direct march to the truth, science advances in a zigzag path that involves many false starts, detours, and blind alleys". While certainty remains a panglossian goal, requiring (J) is intended to somewhat tame the vagaries and vicissitudes: within the limits of what one can realistically expect from science, a scientific discoveries ought to be  epistemically robust. (To reiterate once more, which epistemological standards one adopts---precisely what counts as ``realistically expectable", in particular whether one opts for properly historicised standards---must be decided by the historian who writes about a certain discovery.) 

 Finally, the demand (A) that scientific discoveries be epistemically ampliative. It reflects their epistemic \emph{repercussions}: discoveries are supposed to push the boundaries of a scientific field (rather than---to stick to the metaphor---fortify them). (A) encapsulates the intuition that scientific discoveries propel progress.\footnote{We use ``progress” in a permissive, ecumenical sense, with no specific commitment to any position in the present debate (see e.g. Niiniluoto, 2019; Dellsén, 2019). Here, we remain pluralists about progress.} 
 
 A scientific result R's epistemically ampliative effect consists in at least one of the following. R ... 
 
 \begin{enumerate}
     \item[(A1)] ... extends or broadens; or
     \item [(A2)] ... deepens; or
     \item[(A3)] ... contradicts or rectifies/revises; or
     \item[(A4)] ... adds something completely new and unprecedented to 
 \end{enumerate}

our existing knowledge or understanding.

Our scientific knowledge or understanding is extended/broadened, in the sense of (A1), if a result enlarges them: new elements are (cumulatively) added---for instance, when one finds a novel application (explanation or prediction) for a theory, or one spots a new bug species. 

Our scientific knowledge or understanding is deepened, in the sense of (A2), if we find a deeper explanation or description for a phenomenon (say, a causal mechanism), or if a more fundamental theory is propounded that supersedes (and reduces to) its predecessor and retains the latter's earlier successes. 

An epistemically ampliative R can also conflict with received understanding and knowledge, as per (A3). In this case, R reveals a defect or shortcoming. Anomalies that clash with what what our background knowledge would have us expect are cases in point. Often such results can also correct existing misconceptions. For instance, a host of diverse phenomena forced us to drastically revise our assumptions about the universe's matter content, and allow for Dark Matter. 

Lastly, (A4) deals with epistemic amplification of a more radical hue: R can also be something unprecedented, with little to no continuity (nor \emph{obvious} conflict) with prior ideas or knowledge. Cases in point are pulsars, whose regular, and extremely fast rotation period puzzled astronomers, or prions (i.e. misfolded proteins, causing e.g.\ mad cow disease), which could propagate without nucleic acids and cause.

(A1)-(A4) express the sense of advancing science, the core Kuhnian and Popperian thought about discoveries that our model seeks to salvage: discoveries are problems, or solutions to problems, that drive progressive change. Hence our account's appellation.


Let's wrap up our preliminary presentation of the change-driver model's conditions for discovery, by inspecting scientific results for which the conditions (P), (J), and (A) fail to hold. This will illustrate two points. First, it lends further plausibility to our proposal to regard them as individually necessary and jointly sufficient criteria for a discovery. Secondly, by showing what scientific results are excluded, we show how stringent our account's requirements on a discovery are.  

Consider first situations where a new concept has been introduced, such as energy, entropy, or inertia in Newton's First Law. \emph{On their own}, they neither pose nor answer a question in a sufficiently constrained context: they violate (P). They are best viewed, to borrow Einstein's famous phrase, as ``free creations of the human mind": rather than discoveries that we stumble on in some more or less narrowly circumscribed search space, we \emph{invent} them. 

This different classification aligns with how we appraise their adequacy: we judge them by their fertility in scientific practice, on the basis of whether they can profitably be \emph{exploited}. For this reason, they also contravene (A): whether they ampliatively impact science depends on proven utility; intrinsically, they don't constitute advances. 

Scientific models, such as the Hot Big Bang model or the BCS theory of super-conductivity, don't qualify as discoveries \emph{per se}, either. Of course, they can \emph{lead} to discoveries: the Cosmic Microwave Background, and an energy gap in the Fermi surface, respectively, in the two examples. Models, however, are typically underdetermined by theoretical or observational constraints. Their construction involves creativity and skills that don't easily translate into tangible epistemic constraints (see e.g. Bailer-Jones, 2009). Similar remarks as in the case of concepts apply to their appraisal of adequacy. Moreover (and occasionally even independently of) their utility, we indeed prize scientific models and novel conceptual apparatus for their \emph{creativity} or \emph{adroitness}---virtues that can be, and often are, orthogonal to actually advancing science.   

The case of scientific models---ubiquitous in, and of paramount significance for, science---provides an apt opportunity for clarification: by barring certain items from the category of ``discovery" one doesn't denigrate their scientific value, neither intrinsic nor instrumental. Models often exemplify the advancement of science: they are indispensable \emph{vehicles} for gaining understanding, and extracting successful explanations and predictions. Specifically, as far as (P) is concerned, problem-solving/posing in a tightly constrained search space is only \emph{one} kind of scientific activity; not all bona fide research is of this kind. Scientific innovation doesn't exclusively come about through discoveries, a process that, with respect to their involvement of constraints, resembles the solving and posing of puzzles. Also more creative, less constrained processes engender scientific innovation.

A third group of scientific results that don't satisfy (P) are new interpretations\footnote{For our purposes, we'll gloss over the debate over whether to regard these as distinct theories, or merely notational variants of each other. Our sympathies for the former option notwithstanding, our main point here is that they arguably violate (P).} of scientific theories. The multitude of interpretations of Quantum Mechanics are cases in point, but also Einstein's special-relativistic kinematics (Janssen, 2002b), or alternative formulations of Classical Mechanics (say, Hamilton-Jacobi theory or Barbour-Bertotti theory). These interpretations are \emph{options} which are typically not compellingly forced on us. They typically aren't sufficiently constrained by (robustly accepted) background assumptions to qualify as problem solutions in our more narrow sense. Again, this is neither to impugn their legitimacy or value nor to deny that choices amongst them can have a rational---but perhaps more philosophical---basis. 

Akin are the cases of (re-)axiomatisations of known results (say, Carathéodory's 1906 axiomatisation of thermodynamics), or taxonomies (such classification systems for stellar types, or the Linnaen system for organisms). They violate (P) because they are primarily concerned with a different question: whereas problems inquire into the connection of a datum with a wider cognitive-intellectual context, (re-)axiomatisations and taxonomies systematise or (re-)structure a \emph{given} array of items. The quest for accounting for a datum---as characteristic for a problem---means to place it in a more comprehensive nexus; taxonomies and (re-)axiomatisations, by contrast, are primarily concerned with internal order. Needless to say, both (re-)axiomatisations and taxonomies often purvey substantive understanding (in the sense of, for instance, Elgin, 2017). Moreover, they often \emph{enable} discoveries---as in the case of Mendeleev's periodic table (see e.g.\ Schindler, 2018, Ch.3.5---even though, on our account, they don't constitute discoveries themselves.    

Examples of scientific results at odds with (J) abound. At present, according to our account, it would be wrong---at best, premature---to speak of the discovery of multiverses in cosmology, tachyonic particles, large extra dimensions, or supersymmetry in high-energy physics, even though, \emph{if true}, these speculations would momentously advance science. 

Again, the preclusion of such speculations doesn't imply diminution of their scientific value (see e.g.\ Achinstein, 2018 for a systematic defence). In particular, Laudan (1977, pp.108, see also Nickles, 2006) has drawn attention to the distinction between pursuing and accepting a theory: criteria for assessing acceptability or the epistemic adequacy (truth, explanatory power, empirical accuracy, etc.) typically differ from criteria for gauging which promising---but still inchoate---fledgling ideas deserve further scrutiny for further development. Speculation can be fertile or sterile. The change-driver model's refusal to elevate intriguing ideas that violate (J) to discoveries merely insists on distinguishing between \emph{would-be} discoveries, and actual ones---a distinction paralleling that between reasonably hoped-for and accomplished success (in, say, passing an exam);  ``discovery" is a term signalling \emph{actual} epistemic credentials (cf. Hooper, 2024 for a similar emphasis).  

Another prominent group of results typically viewed as flouting (J) comprises scientifically obsolete ideas. Unless adopting an anti-realist view---an \emph{option}, but not one that our account is wedded to---many user the notion of a discovery would shy away from including cases such as phlogiston, the caloric, or the luminiferous ether as discoveries---even though these entities were once scientifically reputable, and, if true, would have plausibly satisfied (P) and (A). 

With the benefit of hindsight, present-day historians have overwhelming reasons to regard them as non-referring. To our mind, historians are free (but also not obligated) to use contemporary knowledge \emph{cum grano salis}, where and how they see it fit (see e.g.\ Oreskes, 2013 and Chang, 2009). If they plump for adopting today's epistemic-evidential standards as a selection principle for what can be discovered, phlogiston, caloric or the ether are out, on account of violating (J). 

Lastly, which scientific results does (A) rule out? Three common groups spring to mind. The first contains elements that fall into a more didactic or pedagogical genre: say, textbooks or review articles (provided they don't contain novel material, of course). Obviously they can be hugely influential and promote science; scientifically, only they---rather than the original results they present---may fully unlock an idea's potential for further inquiry in a way that facilitates use of, and inspires, other researchers (resulting in further discoveries). Nonetheless, on their own, they don't constitute advances in science. The same remarks apply to other works aiming at popularisation and dissemination (such as Dawkins' ``Selfish Gene" idea).  

A second group of bona fide scientific results that usually don't meet (A) are replication studies (see Matarese, 2022; Matarese \& McCoy, 2024, with whose taxonomy of replication we broadly concur). Their results bear on the assessment of systematic or statistical errors in earlier findings. Albeit without question of great importance in science, their function is verification or consolidation, rather than epistemic amplification in the sense of (A). They probe, and try to ameliorate, the epistemic status of scientific results: for instance, by increasing confidence levels. 

A third and final set of results slips through the cracks of (A) as a matter of historiographical perspective. A fuller discussion of the latter point will be deferred to later on. Here, we merely observe that every historical narrative inevitably operates at a certain level of descriptive grain: one must choose a minimal resolution threshold for scientific events, below which one decides to gloss over them. Once a historian has adopted such a level of historical-descriptive resolution, certain scientific results will no longer count as scientific advances in the sense of (A) (even if for a finer resolution, they plausibly might). The significance of scientific advances lies on a continuum. Typically, however, historians have good reasons---depending on their historiographical focus---to introduce a cut-off for what to consider. This cut-off affects what counts as a scientific discovery (on our model, qua satisfaction of (A)), and what the historian will set aside as making too marginal a contribution.

Generic histories of astronomy, for instance, will pass over the observation of an otherwise unspectacular object in the Kuiper belt during the 2010s. A more fine-grained history for the specialist, however, might well enumerate it as a small discovery. By the same token, generic histories of particle physics pass over the determination of the 76th decimal of the fine-structure constant. Such scientific results consequently don't qualify as discoveries. The exclusion is owed to the historiographical context: the historian's vantage point and interest of inquiry. 

This concludes our presentation of the change-driver's gist. Next, we'll further refine the idea, with a more fine-grained taxonomy of discoveries. It allows historians to individuate particular discoveries, and to categorise them, in ways that track their epistemic/justificatory status, and their role for the dynamics of science.

\subsection{A more fine-grained taxonomy: theoretical, empirical-physical and synthetic discoveries.}

In the foregoing, we defined a discovery in terms of warranted, epistemically ampliative problems or their solutions. Here, we'll put forward a finer distinction amongst discoveries---depending on whether their objects arise in a more theoretical-conceptual or in a more empirical context, or whether the discovery bridges the two contexts.

Consider the discovery of the positron. Thus stated, one provokes a request for disambiguation: what \emph{exactly} are we asked to contemplate---Dirac's (1928/1929) prediction of positrons or Anderson's (1932) first detection (see e.g.\ Wilczek, 2004 for details)? In an equally legitimate sense, both plausibly count as distinct individual discoveries. 

How to dispel the ambiguity? Dovetailing with our characterisation of discoveries in terms of problems, a straightforward terminological clarification suggests itself. The cases of Dirac and Anderson essentially differ over the datum associated with the discovery qua problem. For Dirac, the datum was an apparently positively charged solution he found to his equation of the electron. ``How to account for this positively charged electron solution?" was Dirac's problem question. His answer consisted in the mathematical structure of his electron theory, together with his proposed physical interpretation of it (Dirac's ``hole theory")---a surprising consequence of what, the \emph{theory asserts}, should exist. The answer was incomplete, however. It remained silent on whether such a solution actually corresponds to something real. Dirac's problem solution fell short of interlocking with the more empirical-factual parts of our scientific knowledge. Anderson’s datum, by contradistinction, was observational data: visible tracks of cosmic rays in a cloud chamber. He accounted for the data by interpret it as evidence for a particle \emph{like} an electron, but with opposite sign of charge. Thereby the datum---a piece of empirical-factual knowledge---was linked with the more theoretical parts of our background knowledge, even though only at a superficial level (without deeper theoretical foundations).   

It stands to reason that distinct data prompt distinct questions; they call for distinct answers. It therefore seems natural to chalk Dirac and Anderson up with distinct problems---and hence, on our account, with distinct discoveries.     

These reflections suggest a natural, sufficient criterion for individuating discoveries. (Necessary criteria needn't detain us, see however Nickles, 1981, pp.97; Meheus \& Batens, 1996, pp.173.) Two discoveries are distinct, if they are concerned with different \emph{topics}---if, that is, as problems (or problem solutions), they deal with distinct problem questions, generated by different data. This even holds when the \emph{same} solution accounts for both questions. Such a situation can naturally be treated as the discovery of a novel aspect, or consequence, of the earlier discovery (which accounted merely for the original datum), and as such a new discovery in its own right. Alternatively, if both solutions were found simultaneously, one may regard the event as the discovery of an underlying common origin in the sense of e.g.\ Janssen (2002a)  (assuming, of course, that the other criteria for a discovery are satisfied).

Against this backdrop, let's introduce a more refined taxonomy of discoveries. It's often expedient to categorise a problem-generating datum $X$, according to where on the theoretical/observational spectrum it falls.\footnote{Insofar as, following e.g.\ Quine (1951), one disputes the analytic/synthetic distinction, a clear-cut distinction between empirical and theoretical distinction is likewise blurred at a fundamental level. Typically, though, in specific historical/scientific contexts at least pragmatically it can be fruitfully upheld.} For Dirac's discovery, the datum arose from primarily theoretical considerations: the mathematical analysis and interpretation of his electron theory. For Anderson's discovery, the datum arose from his analysis of empirical data. 

As we stressed, problems trigger the process of integrating the datum that generates them (as well as our proposals for accounting for it) into a wider scientific nexus. They instigate the search for means to forge and strengthen links to other parts of our knowledge and understanding. Where on the theoretical/observational spectrum a datum falls determines the kind of links that satisfactorily accounting for $X$ requires: typically, for a more theoretical datum links to more observational parts of our web of beliefs are especially coveted (and vice versa). In the same vein, this also determines what satisfactory justification for a problem-solution requires.        

On the basis of categorising their problem-generating datum $X$, we can thus group discoveries into three broad types:
\begin{enumerate}
    \item \textbf{\emph{Empirical-physical} discovery.}\\ The datum $X$ belongs more to the empirical-physical realm. The object of an empirical-physical discovery concerns the empirical-physical reality of a phenomenon. It's referred to by the datum $X$ itself; alternatively (as a different convention of designation), by a (possibly even heavily) theory-laden term. 
    \enumeratext{Archetypal empirical-physical discoveries are: various phenomenological laws (e.g. the Snell-Descartes-ibn-Sahl law of refraction, or the Balmer series in atomic physics), place cells, or weight gain in chemical substances after combustion---as is, with its merely rudimentary conceptualisation, Anderson's discovery of the positron. A spectacular example of our own times concerns Dark Matter (see Martens, 2021). A multitude of robust empirical phenomena is known---empirically-materially discovered by, inter alios, Kapteyn, Oort and Zwicky (see Bertone \& Hooper, 2018). A plethora of possible models has been mooted to account for them. To date, none has accrued conclusive evidence.
    
    For its full integration into our wider body of scientific belief, we must put an empirical-physical discovery in contact with more theoretical knowledge and understanding: it stimulates the search for links to---already established, more speculative, or still to be conceived---theoretical parts of science.} 
    
    \item \textbf{\emph{Theoretical} discovery.}\\ The object of theoretically discovering\footnote{Note that scientists themselves use this term (e.g.\ Wilczek, 2004, p.68).} something is to discover it qua \emph{theoretical} result---that is, as treated, and originating within a given theoretical context. The datum $X$ that generates the problem associated with the discovery occupies a spot on the more theoretical part on the theoretical/observational spectrum.
    \enumeratext{Paradigmatic theoretical discoveries include asymptotic freedom in QCD, gravitational waves or singularity theorems as theoretical-mathematical facts within General Relativity, or the Bell inequalities---but also inconsistencies within or between theoretical frameworks (e.g.\ the ultraviolet catastrophe of classical-electromagnetic blackbody radiation, or the Gibbs paradox, which revealed a problem in the foundations of traditional thermodynamics). 
    
    Note that due to the  relativisation to a particular theoretical context, talk of theoretical discoveries remains meaningful even if one rejects that context as a scientifically inadequate theoretical description. 
    
    In the \emph{empirical} sciences, it's vital, of course, that theoretical discoveries latch onto something outside of their theoretical context.\footnote{In the case of logico-mathematical discoveries, such as the discovery of fractal geometries, Lie groups, or complex numbers, the very notion of a discovery requires formulating a mathematical theory (Bartels, 2021, Ch.3.9), but not a connection to empirical-observational parts of our knowledge.} For their full integration into our wider body of scientific belief, one must thus put them in contact with more empirically grounded knowledge: theoretical discoveries stimulate the search for links to---already established or still to be found---empirical parts of science. Here, synthetic discoveries step into the breach.}

    \item \textbf{\emph{Synthetic} discoveries.}\\A synthetic discovery synthesises aspects of theoretical and empirical-physical discoveries. Its object is a relation between a theoretical concept/conceptualisation and the world: the perspicuous combination of theoretical-conceptual resources, and empirical parts of our knowledge. To synthetically discover something means to discover that, as a theoretical idea, it can be successfully (or, against expectations, \emph{un}successfully) used to explain, model or accommodate certain phenomena.  
\end{enumerate}

Synthetic discoveries often take the form of corroborations (e.g. Brownian motion as a confirmation of the kinetic theory of heat, or Frisch’s deciphering of the bees’ waggle-dance) or falsification (e.g. the detection of neutrino oscillations, refuting the original assumption of the neutrino’s masslessness).\footnote{To deflect the suspicion that one here succumbs to realist temptations, we hasten to add that a synthetic discovery can be agnostic with respect to realism: the metaphysical quarrel between realists and anti-realists seldom stymies agreement on which theory to regard as the most explanatory or empirically best-supported (e.g. the standard model of particle physics, or the modern evolutionary synthesis). Synthetic discoveries concern such theory appraisals.} 


The three types of discoveries clearly differ with respect to the kinds of problems associated with them. The resulting taxonomy cuts natural distinctions. The history of the electron exemplifies this (cf. Norton, 2000; Bains \& Norton, 2001; Wilczek, 2004). A polysemous umbrella term, the ``discovery of the electron" encompasses empirical-physical discoveries (e.g. the results of Millikan’s oil drop-experiment establishing the existence of an elementary unit of charge, or the photoelectric findings of Lénárd’s cathode ray experiments), theoretical ones (e.g. the natural incorporation of spin via Dirac’s theory of the electron) and evidential-explanatory accomplishments constituting triumphant synthetic discoveries (e.g. the stunningly accurate prediction of the electron’s gyromagnetic ratio in quantum electrodynamics).\par

Vis-à-vis our account's emphasis on the dynamics of science, it's fitting to close by highlighting difference between theoretical and empirical-physical discoveries, on the one hand, and synthetic ones, on the other. The former ``itch": how to link them to empirical-observational and theoretical parts of our knowledge, respectively? Both kinds of discovery have an inherently dialectical moment; they call for more. Synthetic discoveries answer that call: they build the sought bridge between theoretical and empirical-observational parts of our knowledge. Thereby, synthetic discoveries contribute to a theoretical discovery's epistemic credentials. Through such boosted consolidation, it's gradually canonised to an established part of one's background knowledge (and can itself serve as a constraint on other problems and their solutions).


\subsection{Merits}
Here, we'll discuss the merits of our account. It overcomes the flaws of the received models (\textbf{§3.3.1}), whilst preserving their most convincing advantages (\textbf{§3.3.2}). It furthermore proves fertile in shedding light on a number of salient aspects of discoveries, related to their communal dimension (\textbf{§3.3.3}).\par

\subsubsection{Remedies to other models' defects}
How does our account avoid the four principal shortcomings, diagnosed earlier (\textbf{§2.4}): the restriction to material objects, reticence about the link with the dynamics of science, the pro-realist bias, and the Problem of Theoretical Continuity?  

In \textbf{§3.1}, we already touched on the first: our account imposes no restrictions on the datum that generates a problem, and consequently on the possible entities that can be discovered.

Secondly, in our view, the connection with the dynamics of science is essential for discoveries. We therefore regard the extant accounts' opaqueness with respect to that connection not just as a shortcoming, but a grave defect. By marked contrast, the change-driver account builds it into the notion of a discovery ab initio with its advancement clause (A): discoveries challenge our existing body of knowledge and understanding, and/or proffer resolutions of these challenges---resolutions which, in turn, increase our knowledge and understanding. \emph{That} problems and the struggle for solutions are decisive factors in the dynamics of science is hardly controversial (e.g.\ Laudan, 1977; Popper, 1994; Shan, 2023).

The change-driver model addresses the third key defect of the other post-Kuhnian models (without swinging to the anti-realist opposite extreme, as Kuhn espoused it). It divorces the notion of a scientific discovery from the realism/anti-realism debate. Not relying on substantive premises on either side of the debate, our account stays neutral. 


Historians are free to reserve the notion of ``discovery" to those problems and solutions that are still accepted. Should they so wish, they merely have to add suitable selection principles as additional constraints. But they are no less free to relinquish realist aspirations, or any Whiggish (presentist) proclivities. They may apply the notion of a discovery in a fully historicised manner---as historical agents would have used it, in light of \emph{their} cognitive-intellectual context. Our account's explication of ``discovery" is available to realists and anti-realists alike. To be sure, the resulting historiographical narratives will differ. That is hardly an objection, however; we rather embrace the resulting historiographical pluralism. Judgements about which historical analyses are more illuminating are matters we leave to historians.

Compatible with a pluralist approach to progress, our account isn't tied either to any specific position regarding notions of progress.\footnote{It’s natural, but by no means inevitable, to adopt Laudan’s (1977, 1996, Ch.4) view of progress as problem-solving (see also Shan, 2023).} In particular, the model isn’t tied to a realist construal of progress in terms of verisimilitude (increasing truth-likeness).\par

This leads to the fourth problem besetting the standard accounts. The ``Problem of Theoretical Continuity" doesn’t arise in the change-driver model. Discoveries exist in specific \emph{historical} problem contexts; they inextricably belong to (historically situated) systems of knowledge and understanding.\footnote{Caneva (2001, 2005) underscores that a crucial task of the historian is to faithfully reconstruct this place of a discovery in a problem situation: ``[...] what a scientist is typically credited with having discovered often differs significantly from the way in which the scientist himself characterized his work" (Caneva, 2001, p.1).} The relation to later problem contexts is, of course, instructive. By tracing the evolution of problem contexts, historians help us understand the history of a discipline. Sometimes, continuity between past and present science can be discerned (see e.g. Bain \& Norton, 2002 for the case of the electron); more often than not, though, their relationship is loose and motley (Chang, 2003).

For judging \emph{whether} a problem and/or its solution counts as a discovery, questions of continuity of problem contexts are peripheral. (As stressed, historians may decide to employ sufficient closeness to contemporary knowledge and epistemic criteria as additional selection criteria for discoveries. But this is an option, not a principled feature of our account.) The need lapses to postulate a strong link between the scientific discovery in its historical understanding, and the discovery as \emph{we} interpret it.

The change-driver model is historically sensitive. To classify a scientific result as a discovery doesn’t commit us to anachronisms. A result can be granted that status on terms natural to its historical problem context. This allows us---if so inclined---to honour past theories' achievements. The ether theory, for instance, entailed several genuine empirical-physical discoveries (see e.g. Janssen \& Stachel, 2004). In the same vein, we can recognise the various explanatory and evidential successes of now-defunct, but formerly rightly cherished theories, such as the caloric theory (see e.g. Chang, 2003) or the Steady-State Theory (see Kragh, 1996), as synthetic discoveries.\par

\subsubsection{A unified account of what-that and that-what discoveries}

The change-driver model provides a unified account of ``that-what” and ``what-that”-discoveries. An empirical-physical discovery, which precedes its satisfactory conceptualisation and theoretical grasp, fits the bill of Kuhn's ``that-what" discoveries. In the same vein, \emph{if} a theoretical framework is epistemically well-established, and a new synthetic discovery is made whose empirical counterpart is subsequently verified, we obtain a ``what-that"-discovery, with its attendant ``material demonstration". For these special cases, we recover the---convincing---gist of Kuhn's model: his distinction between those two types of discoveries. 

The change-driver account covers, however, a broader scope of discoveries. It recognises also more purely theoretical, or more purely empirical discoveries as discoveries in their own right---in line with common practice and parlance. 

A crucial advantage of the change-driver account lies in its extrication from Kuhn's problematic two-stage model for the dynamics of science. As elaborated in \textbf{§3.2}, we needn't invoke paradigms and their cataclysmic shifts to appreciate the impetus---for Kuhn, a paradigm-enforced compulsion---to complement more theoretical discoveries by more empirical-physical ones, and vice versa. For a scientific result to acquire and unfold its scientific significance, it must be placed in, and ideally cohere with, our wider web of beliefs (see Blackwell, 1969, esp. pp. 140 for details; also Bonjour, 1985; Elgin, 2017, esp. Ch.1-4 for in-depth studies of the broader epistemological point). Through this integration, our scientific understanding grows---one of the cognitive aims of science.

We try to make sense of a discovery in light of our prior knowledge, and vice versa, with cross-checking and adjustments until equilibriation (see Turner, 2018, who stresses precisely that dialectic for the case of cosmology).\par


\par


\subsubsection{Shedding light on the communal dimension of discoveries}

With its emphasis on advancing \emph{science}, the change-driver model has ab origine a communal dimension. Science requires a scholarly community within which it operates---another profound Kuhnian lesson (Longino, 1990 and Solomon, 2001). Scientific discoveries, qua aspects of science, can’t happen in isolation (see also Michel, 2022, sect.2.5); they must conform to the organisational and normative structures of science as a community (see e.g.\ Ziman, 2002). (This directly demarcates commonplace from scientific discoveries: rather than because of psychological-subjective familiarity---as in Hudson's model (\textbf{§2.2})---a marten in the attic is a finding, disconnected from the (any) scientific community, and thus no scientific discovery.)\par

This communal dimension \emph{explains} a couple of phenomena that, on the other models, look puzzling, or are consigned to white noise of human hustle and bustle in science. They are traditionally banished to the ``external history" of science, not susceptible to further rational analysis—and maybe “of great interest to empirical psychology; but [...] irrelevant to the logical analysis of scientific knowledge” (Popper, 2002b, pp.7). By contrast, to our minds, these communal aspects are central to scientific discoveries (cf.\ M\"unster, 2022, sect. 4.8). An adequate explication of the notion must account for them.

First, recall one of our quibbles with Hudson’s ``novelty condition" (\textbf{§2.2}): which social group is supposed to be relevant for judgements of temporal priority? In light of the stunning scientific achievements of, say, Chinese science and the Islamic world the question becomes pressing (with unacceptable eurocentrism looming). Hudson’s model ended up with an outr$\acute{e}$ answer; the other models skirted the question altogether. The change-driver model, with its explicit community-relativity, embraces a pluralism of communities: the notion of a scientific discovery is always relative to a scientific community. The latter, in turn, is a complex system, a local network of inquirers with strong information links and an institutionalised form of critical exchange (see e.g.\ Bunge, 2003), rather than to any \emph{arbitrarily selected} social group, or (pace Hudson) to ``all Earth-bound rational beings over the course of all time”.\footnote{Insofar as science has---fortunately!---always been global (see e.g. Poskett, 2022), any talk of local scientific communities remains, to some extent, an idealisation that belies the more complex historical reality; ideas and knowledge have always travelled across borders. Yet, this idealisation underlying orthodox historiography of science has, we think, value in at least a rough-and-ready sense. It’s beyond the present paper’s ambit to question it.}\par

The second advantage is also related to another blemish in Hudson's account. It concerns the priority condition. Discoveries (and discoverers, for that matter) are usually assumed to fulfil it. But why demand that a discoverer be the first to make the discovery? Rather than stipulating by fiat, the change-driver model can, to some extent, \emph{derive} it. \par

According to our account, temporal priority is, strictly speaking, inessential for a discovery. What matters instead is that the result increases knowledge and usually, that it sparks off further developments. More often than not, this is going to be \emph{correlated} with temporal priority. Nonetheless, the correlation doesn't always hold: unpublished work or publication in obscure journals may later be found to have presaged a claimed scientific discovery. We bite the bullet that adverse circumstances, or fatal decisions on the side of the researcher, can prevent such \emph{private} discoveries from becoming \emph{scientific} ones (or modally paraphrased: they can prevent \emph{potential} scientific discoveries from becoming \emph{actual} ones). Luck, good and bad, is an ineliminable factor in actual scientific discoveries (see Copeland, 2018). Any realistic model of discovery must allow for this.\par

A third phenomenon that our account effortlessly explains concerns the hierarchy of discoveries:
why \emph{aren’t} all scientific discoveries celebrated equally? In particular, why does the fame associated with a scientific discovery seem to be correlated with the scientific impact it makes? The first discovered exoplanet, for instance, made a big splash (see e.g. Winn, 2023). Now discoveries of exoplanets are a dime-a-dozen; the euphoric reaction to discovering a new one has abated. The change-driver model easily explains this in a science-internal manner, vindicating the rationality underlying the social practice: a discovery is, in the main, celebrated according to its scientific value, its significance. The first exoplanet discovered validated a range of observational techniques, instrumentation, theory; determining its properties like mass, radius, composition, etc presented new problems to solve. By the umptieth exoplanet, the flurry of activity following such a discovery has subsided; in terms of significance, scientific results of the same kind have diminishing returns. On the change-driver model, their status as discoveries pales—and concomitantly the occasion for accolades. For the other accounts, the discovery of the first and million\textsuperscript{th} exoplanet are both discoveries. Nothing more, it seems, can be said---nothing in particular about the rewards meted out to the discoverer. \par

Fourthly, let’s wonder: why are scientific discoveries celebrated \emph{at all}? What makes them so nigh-universally appreciated and valuable? Bracketing all normative issues, the extant accounts remain silent on that. The change-driver model, by contrast, gives a simple answer, similar to Kuhn's: discoveries---epistemically ampliative forms of problem-solving or posing---realise the cognitive aims of science; they are natural ``units of scientific progress and are accordingly valued" (Arabatzis, 2006, p.220). 

The point generalises to the importance attached to scientific discoveries, visible at the social level: in priorities disputes, and the general reward system for discoveries (prestige, prizes, and eponymy in particular). These practices implement the ``normative structure of science" (Merton, 1942; see also Ziman, 2002), the ``complex of values and norms which is held to be binding on the man of science" (Merton, 1973, pp.268). One norm is originality, or the novelty of insights. A premium on a scientific result’s originality incentivises researchers to strive for innovation. In order to advance science, however, novel ideas in science must be accompanied by ``communalism": they must be made publicly available. Results can only unfold their potential for scientific progress if shared in the scientific community (not least for further critical scrutiny). Both norms, originality and communalism, thus explain the credit/recognition system for discoveries in a science-\emph{internal} manner: those practices reflect the norms inherent to science; through their institutionalised implementation, they promote scientific progress. The change-driver model hence is not only tied to the social practices surrounding scientific discoveries. It also anchors them and their rationality in science itself and its ethos (see Merton, 1957 for a detailed analysis along those lines): \emph{given} the institutional values and norms of science, scientific discoveries are indeed worthy of applause.\par

\section{Illustration: the discovery of the expanding universe}

To show off our model in action, we will apply it to “one of the major milestones in the development of the science of astronomy during the last 100 years, [...] one of the founding pillars of modern cosmology” (IAU, 2018): that the universe expands. Who—if anyone—made this discovery remains contentious. Why choose this case study over a more prototypical and widely analysed case of scientific discovery like the discovery of oxygen? We feel that the discovery of the expanding universe offers a variety of advantages for our purposes. First, it involves, over a relatively short period, empirical-physical, theoretical, and synthetic discoveries. It is an interesting, philosophically rich case that has received too little attention in the philosophical literature. Second, unlike older discoveries like that of oxygen, the historical record is fairly complete and uncontested (Musgrave 1976;  Hoyningen-Huene 2008; Chang 2010). Not only are published papers readily available for analysis, so are draft papers, letters, speeches, conference proceedings, notes, and so on.\footnote{It is worth stressing how useful it is to have access to all these documents. Discoveries, on the change-driver model, must be properly justified. In order to avoid judging the past solely by the present standards, we wish to have some sense of what the evidential standards of the time were. While today the influence of a paper could, in part, be assessed through (say) citation count, citation was rarer during the period of interest. However, letters and other communiqués reveal what work was having an impact.} Further historians may quibble over who to label as a discoverer in this case, but they do not quibble over the \emph{basic facts} of the case- who did what and when. After reviewing the main historical developments in \textbf{§4.1},\footnote{We follow the presentations in North (1965, Ch.5-7); Smith (1983, Ch.5); Ellis (1989); Kragh (2007, Ch.3), and Nussbaumer \& Bieri (2009), to which we refer the interested reader for details.} we’ll compare the verdicts of the extant accounts of discovery for the case of the expanding universe in \textbf{§4.2}. Finally, \textbf{§4.3} will apply the change-driver model’s perspective. While we will identify individual discoverers, it is important to note that the change-driver account provides a \emph{model} of scientific discovery, not an algorithm for identifying discoverers. But identifying discoverers is an important part of presenting historical and pedagogical accounts of scientific episodes, and this identification is, of course, impossible without some sort of account of discovery.

\subsection{Historical overview}

The gates to modern theoretical cosmology were flung open in 1917, when—less than two years after completing his new theory of gravity, general relativity (GR)—Einstein applied it to the universe as a whole (see e.g. O'Raifeartaigh et al., 2017 for details). This was the first cosmological model, an exact solution of GR’s field equations for the cosmos at large. In it, Einstein assumed, matter was distributed uniformly (i.e. homogeneously and isotropically), and the universe as a whole finite (i.e. its spatial geometry closed). Einstein’s model quickly stimulated further investigations—the pioneering work of modern cosmology’s founding fathers (amongst them, Friedmann, de Sitter, Eddington, and Lemaître, about whom we’ll hear more momentarily). \par

Two peculiarities of Einstein’s result are worth pointing to. First, Einstein’s primary motivation wasn’t to apply GR to a new domain—one where one might perhaps expect incisive discrepancies with Newtonian theory. He didn’t intend his model as a realistic description of the universe, apt for empirical tests (nor was Einstein even particularly informed about the state of the art in astronomy). “Instead, Einstein’s foray into cosmology was a final attempt to guarantee that a version of ‘Mach’s Principle’ holds” (Smeenk, 2013, p.228), i.e. the idea that inertia is fully determined by matter. In that regard, de Sitter’s work soon dashed Einstein’s hopes. \par

Secondly, and of key relevance for our purposes, Einstein’s model was static: the universe it represented stays the same across time. To balance out the gravitational attraction of matter, and thereby to achieve staticity, Einstein had to modify his original field equations of 1915—by including the cosmological constant term $\Lambda$ (introducing a free parameter). “Although he originally treated this as only a simplifying assumption, Einstein later brandished the requirement that any reasonable solution must be static to rule out an anti-Machian cosmological model discovered by de Sitter. Thus [...] Einstein was blind to the more dramatic result that his new gravitational theory naturally leads to dynamical models. Even when expanding universe models had been described by Alexander Friedmann and Georges Lemaître, Einstein rejected them as physically unreasonable” (op.cit., p.229). In clinging to the static nature of the universe Einstein conformed to the prevailing consensus, rooted in the time-honoured doctrine of the immutability of the heavens\footnote{It’s imperative to appreciate how different the received wisdom about the universe was in 1917: recall that Einstein was writing two years before “The Great Debate” over the size of the Milky Way and the existence of “nebulae”, i.e. extragalactic systems (see e.g. Smith, 1983, Ch.1-3)!}---and a tenuous observational basis in “the small velocities of stars” (Einstein, 1917, p.139).   \par

Throughout 1916 and 1917, de Sitter propounded his own model of a homogeneous and isotropic universe (de Sitter 1916a; 1916b; 1917). In contrast to Einstein’s model, it was devoid of matter; the model contained only a cosmological constant $\Lambda$. Likewise in contrast to Einstein, de Sitter “took his model seriously enough to study its observational consequences, as we will see” (Smeenk, 2013, p.244). \par

Over the next decade, Einstein’s and de Sitter’s models remained the only well-known relativistic models of the universe; most cosmological investigations focused on those two rivalling models. “The major conceptual innovation introduced by Einstein was the very possibility of a mathematical description of the universe as a whole, but it was not immediately clear what observational and physical content these abstract models possessed” (op.cit., pp.242).\par

De Sitter makes two points important for our story. First, unlike Einstein, he countenanced a dynamical universe as a physical possibility: “(i)n [de Sitter’s model—in contrast to Einstein’s] [...], if there is more than one material particle these \emph{cannot be at rest}, and if the whole world were filled homogeneously with matter, this could not be at rest without internal pressure or stress” (1917, p.18, our emphasis). This wasn’t quite tantamount to entertaining an expanding universe; the de Sitter model was generally considered a static solution. De Sitter’s spacetime metric didn’t contain a time-dependent scale factor (describing the change of physical spatial distances across time). \footnote{In 1922, however, Lanczos showed how the de Sitter model could, through a change of coordinates, be interpreted as an expanding universe with a (time-dependent) hyperbolic, spatial geometry.\footnote{In his first article on cosmology (1925), Lemaître also introduced new coordinates that rendered the de Sitter metric dynamical (with vanishing spatial curvature, and an exponentially increasing scale factor).} That it actually wasn’t a static model took a surprisingly long time to realise. The waters were muddied by confusion over artefacts due to coordinates (a common ailment of early work in relativistic physics, see e.g. Kennefick, 2007, Ch. 4\&5).} \par

Secondly, de Sitter (seconded, in 1925, by Lemaître) hailed the dynamical character of his model as an avenue for further research (e.g. 1917, p.28): might it, he mulled, explain the receding motion of spiral nebulae? Indeed, de Sitter (1917, sect. 7) predicted a distance-dependent redshift in the spectral lines of far-away objects. Using available data, he even calculated the effect. Moreover, he “took an important step in suggesting that the movements of the spiral nebulae rather than the stars (which Einstein had focused on) should be used as gauges of cosmic structure on the largest scales. The nature of the redshift effect and the precise functional dependence of redshift on distance for the De Sitter model were both matters of substantial controversy for the following decade and a half” (Smeenk, 2013, p.249). De Sitter first erroneously derived a \emph{quadratic} dependence of redshift on distance. Later, Weyl (1923) claimed that if test particles were introduced on geodesics emanating from a common point in the past, objects would appear to recede approximately according to a \emph{linear} redshift-distance relation. Introducing matter in a different manner would lead to different predictions. Improving on Weyl’s result, Robertson (1928) confirmed the linear redshift-distance relation.\par

The prospect of predicting redshifts was enticing. De Sitter’s model “soon became a foundation for further theoretical work, among both astronomers and mathematicians” (Kragh, 2007, p.136)---including Eddington and Weyl. In 1909, Slipher had begun measuring redshifts at the Lowell Observatory, taking the spectra of spiral nebulae. In 1917, he published his results: all 25 spiral nebulae exhibited quite large blueshifts or redshifts. The high radial velocities invited an interpretation as recession velocities, resulting from the (standard) Doppler effect. The redshifts, Slipher averred, may have been the result of the Earth’s motion through space: the nebulae we moved towards would be blueshifted; the ones we moved away from would be redshifted. But by the early 1920s, the redshift observations of spiral nebulae ``left little doubt that there was a systematic recession. [....] (F)rom about 1920 there developed a minor industry based on the [Einstein’s and de Sitter’s respective] models. It was predominantly a mathematical industry” (Kragh, 2007, p.136). The next milestone would be to build a bridge to physics, and astronomical observations in particular: “The question of the relationship between cosmology and the observed redshifts remained unresolved for a decade or so, for other reasons, because it was difficult to distinguish a cosmological redshift (the de Sitter effect) from gravitational redshifts and the Doppler shifts caused by relative motion” (ibid.).\par

For some time, the Einstein and de Sitter cosmologies were the only well-developed cosmological solutions to GR. Two papers, one in 1922 and the other in 1924 (as well as in a semi-popular book of 1923, in Russian), ventured further. In them, Friedmann disclosed a much larger class of solutions: its members corresponded to matter-filled universes that \emph{expanded}. Friedmann procured nothing short of “a complete and systematic analysis of the solutions of Einstein’s cosmological equations that went beyond earlier analysis” (Kragh, 2007, p.141). The Einstein and de Sitter model turned out to be special cases of a richer space of solutions; these solutions demonstrate “the possibility of a world in which the curvature of space is independent of the three spatial coordinates but does depend on time” (Friedmann, 1922, p.377). Specifically, Friedmann had solved Einstein’s field equations of GR for open and closed homogeneous and isotropic universes with time-\emph{dependent} radii of curvature. For different values of $\Lambda$, Friedmann explored the consequences of cyclical models: the universe, in those solutions, periodically expand and contract, undergo accelerated expansion, or contract back into a point. With Friedmann, dynamical cosmological models of GR were squarely on the table—as mathematical possibilities. Friedmann \emph{explicitly} didn’t connect his models with astronomical observations: “(o)ur information is completely insufficient to carry out numerical calculations and to distinguish which world our universe is” (ibid., pp.385-386). No references to astronomical data are included in Friedmann’s papers. For Friedmann, his analysis seems to have been first and foremost a \emph{mathematical} exercise (see also Kragh \& Smith, 2003, pp.146). Did Einstein and others overlook Friedmann’s work? Not at all: he, like most others, simply didn’t attribute to Friedmann’s non-static cosmological models any physical significance. Friedmann had ushered in an “unnoticed revolution” (Kragh, 2007, p.140). Still in 1927, firmly in the grip of the ruling paradigm of a static universe, Einstein unabashedly dismissed non-static cosmological models as, from a physical perspective, “tout $\grave{a}$ fait abominable”. \par

The occasion for Einstein’s aspersion was a meeting with Lemaître. It concerned cosmological models for an expanding universe. They had been the subject of an article by the latter a few months earlier (1927) that “a friend had made (Einstein) read” (Lemaître as quoted in Luminet, 2013, p.1625). Its agenda Lemaître had already foreshadowed two years earlier: “Eddington [1923] writes [...]: ‘It is sometimes urged against de Sitter's world that it becomes non-statical as soon as any matter is inserted in it. But this property is perhaps rather in favor of de Sitter's theory than against it.’ Our treatment evidences this non-static character of de Sitter's world which gives one \emph{possible} interpretation of the mean receding motion of spiral nebulae” (Lemaître, 1925, p.41). Unravelling this thought, Lemaître’s 1927 \emph{opus eximium} “combines the advantages of the Einstein world and the de Sitter world” (Kragh, 2018, p.1334) in order to find a solution to the Einstein field equations that satisfactorily accounts for both the redshifts and the existence of matter. This led Lemaître to a cosmological model with a \emph{time-dependent} radius of curvature—corresponding to an expanding, homogeneous and isotropic closed universe. In the main\footnote{A qualification is due: Lemaître included thermodynamic aspects. “Whereas Friedmann had ignored the radiation pressure, it was an important quantity in Lemaître’s more physical and ambitious project” (op.cit., p. 1338).} Lemaître’s formal results duplicated Friedmann’s; to his work Lemaître was only alerted in the above-mentioned meeting with Einstein. The crucial innovation, as far as the physical thrust of Lemaître’s paper is concerned, was his renunciation of the ruling paradigm of staticity: Lemaître—“in stark contrast to Friedmann” (op.cit., p.1334)—audaciously suggested that \emph{our} universe \emph{in fact} grew in size. “What distinguished Lemaître’s paper over Friedmann’s was in particular his analysis of the galactic redshifts and his pellucid recognition that they were caused solely by the expansion of space” (op.cit., p.1338)—that is, the realisation “that the redshifts were caused not by galaxies moving \emph{through} space, but by galaxies being carried \emph{with} the expanding space” (Kragh, 2007, p.144). \par 

Theoretically, Lemaître derived an approximately linear relationship between the distances and recession velocities of extragalactic nebulae (i.e. galaxies)—in the form still used today (i.e. with the proportionality constant given by the “Hubble parameter”, the ratio between the rate of change of the scale factor and the scale factor itself). He then connected his solution to astronomical observations. Relying on averaged data for the ratio between radial velocities (through data due to Str$\ddot{o}$mberg and Slipher) and distances (through data due to Hubble), Lemaître calculated a rough estimate of the “Hubble parameter”. What, in other words, Lemaître thereby did accomplish was to empirically determine the proportionality coefficient of his theoretically derived linear distance-velocity relationship. He thus made a genuine prediction---that, once sufficiently accurate data were available, they would corroborate that linear relationship (roughly with the proportionality coefficient Lemaître had determined). By contradistinction, what he \emph{didn’t} do was to provide original empirical evidence for that relationship; he couldn't yet verify his prediction. \par

Kragh argues that “Lemaitre’s prediction of an expanding universe made no more impact than did Friedmann’s work. On the contrary, his paper seems to have been almost completely unknown and to have received no citations from other scientists until 1930” (ibid.). The reasons plausibly lie in the relative obscurity of the journal, the \emph{Annales de la Soci$\acute{e}$t$\acute{e}$ Scientifique de Bruxelles}, in which he had decided to publish it (in French).\footnote{Luminet (2013, p.1625) locates the “main obstacle to a larger diffusion of Lemaître’s article” in the fact “that most of the physicists at the time, such as Einstein and Hubble, could not accept the idea of a non-static universe.” According to O'Raifeartaigh (2020, p.9), “it is very plausible that the observational data cited by Lemaître in support of his model were not sufficiently robust to cause Eddington and others to consider expanding cosmologies.”} Through the hands of Eddington, three years after the paper’s appearance, the tide would turn for Lemaître’s fame. But first someone else would step into the limelight: the very person usually, but contentiously, credited with having discovered the expanding universe.  \par

During the mid-1920s, Hubble had ascended to international renown as an astronomer at the world’s most powerful telescope, the Mt. Wilson Observatory. His meticulous observations conclusively showed that observed nebulae were outside of the Milky Way, drastically changing our view of the Universe by populating it with island worlds.\par

Hubble was thus well-prepared for the next problem he tackled—the redshifts of galaxies. As we saw above, both theoretically and observationally, a correlation had been postulated between redshifts and distance of extragalactic nebulae: they appeared to flee from earth the faster, the farther away they were. But no consensus existed as to the functional relationship—or even the reliability of such a systematic correlation between observed redshifts and distances (see also O’Raifeartaigh, 2020).\footnote{Especially noteworthy amongst these early empirical endeavours prior to Hubble was the work of Wirtz (1922, 1924). Despite too insignificant a sample size for further conclusions, he was able to demonstrate (upon suitable grouping of the data) a systematic positive correlation between redshift and distance (and interpreted this as a confirmation of de Sitter’s model), see Duerbeck \& Seitter, 1990 for details. A similar result was achieved by Lundmark (1924, 1925), who drew on Slipher’s velocity measurements and his own distance measurements.
} At a meeting of the Astronomical Union 1928, Hubble probably learnt from conversations with de Sitter and others how his previous work and expertise might be relevant for on-going work in theoretical cosmology. The study Hubble embarked on would, Hubble hoped, also enable a critical test between Einstein’s and de Sitter’s model of the universe.\par

A significant leap forward came with the discovery that Cepheid variable stars can function as `standard candle' distance indicators:“a formidable obstacle had to be overcome before securing a relation between redshift and distance that would be convincing to the majority of astronomers: the accurate determination of the distances to the [spiral galaxies]. But in 1923 and 1924 this barrier was in part removed when Hubble discovered Cepheid variables in nearby spiral nebulae. Before Hubble’s observations of Cepheids, astronomers had been restricted almost entirely to the crude distance indicators provided by the novae, the apparent luminosities and the apparent diameters of extragalactic nebulae” (Smith, 1983, p. 175). 

Cepheids denote a group of stars with a variable brightness, pulsating with a well-defined, stable period. Leavitt (1908) (and Leavitt \& Pickering, 1912) had shown that the length of a Cepheid’s period is correlated with its intrinsic luminosity: the longer the period, the brighter the Cepheid. Together with Hertzsprung’s (1913) calculation of the zero-point luminosity and calibration of the period-luminosity relation (by measuring the distance to Cepheids in the Milky Way through parallax), Cepheids could thus be utilised as so-called standard candles, i.e. as means for determining cosmic distances. Hubble hadn’t observed Cepheids in every galaxy he had redshift data for. (Individual Cepheids are difficult to resolve at high distances, and are not observed in all galaxies.) So he calibrated other assumed standard candles, such as the brightest star in a galaxy or the total brightness of a galaxy, to determine the distance to galaxies without observed Cepheids.\par

Plotting the distances of the galaxies (measured by himself) against their velocities (essentially drawing on measurements by Slipher and Humason), Hubble (1929) “(established) a roughly linear relation between velocities and distances among nebulae for which velocities have been previously published” (p.173). “(M)any commentators have noted that the quality and quantity of the data shown on Hubble’s graph only marginally supported his conclusion of linear relation between redshift and distance for the nebulae […]. However, the graph marked an important turning point […]” (O’Raifeartaigh, 2019, p.9). The data on which Lemaître had relied in 1927, in particular, was contaminated by significant uncertainties due to scatter. The linear redshift-distance relation dimly discernible in Hubble’s data could at best count as preliminary: “the nebular distances were established using a method that was prone to large errors” (op.cit., pp.8). Hubble’s subsequent paper, co-authored with Humason (Hubble \& Humason, 1931) dispelled lingering scepticism. Put in strong terms, “the linear correlation between redshift and distance seemed to be even clearer than before and effectively ended the debate on the existence of a linear relation” (Kragh \& Smith, 2003, p.150).\par

How did Hubble interpret the empirical finding? He never—throughout his life—unambiguously championed its interpretation as evidence of an expanding universe (see op.cit., pp.151 for details). In 1929, Hubble had little to say on the interpretation of his data except that it could be a manifestation of the “de Sitter effect” (p.173): “(i)n the de Sitter cosmology, displacements of the spectra arise from two sources, an apparent slowing down of atomic vibrations and a general tendency of material particles to scatter. […] The relative importance of these two effects should determine the form of the relation between distances and observed velocities” (ibid.). No mention is made of an expanding universe. Also later on, Hubble remained, at best, agnostic about that now-standard interpretation; deferring its interpretation to others (as he expressly wrote in a letter to de Sitter, quoted in e.g. Luminet, 2015, p.5), he was wont to underscore the empirical/phenomenological nature of the linear redshift-distance relation.\footnote{Indeed, that interpretation ran into a chronic embarrassment: the age problem. As implied by that interpretation, given Hubble’s measurements, the cosmos would be younger than the Earth (as inferred from radioactive decay) or the oldest known stars—a glaring absurdity (see also Gr\o n, 2018, p.13).} In line with this, Hubble “was not aware that the proportionality factor between redshift and distance, wrongly named the ‘Hubble constant’ was not a constant since it varies with time” (ibid.)—a variation that follows directly from interpreting the redshift-distance relation in terms of an expanding universe. \par

Hubble’s paper fell on fertile soil. By and large, it had dawned on the astronomy community that Einstein’s and de Sitter’s cosmologies might be inadequate. The question of making sense of the redshift-distance relation became increasingly pressing. At the Royal Astronomical Society in 1930, Eddington went for the jugular: “why (should there be) only two solutions [i.e. Einstein’s and de Sitter’s models, both of which Eddington regarded as static](?) I suppose the trouble is that people look for static solutions” (p.39). Seconded by de Sitter (likewise present at the conference), he enjoined that dynamical solutions be looked into in earnest, as a solution of the redshift conundrum. \par

Reading the reports of this London meeting, Lemaître instantly apprehended that he had \emph{solved} the problem three years earlier in his paper. He immediately wrote to Eddington. Not only was Eddington the leading authority on relativity and astronomy, but he also happened to have been Lemaître’s supervisor during his postdoctoral studies in Cambridge. In fact, a few years before, Lemaître had shared with Eddington a copy of his 1927 paper—which Eddington simply seemed to have forgotten about.\par

In his letter, Lemaître reminded Eddington of his paper. He also attached copies, requesting that they be circulated. Eddington—embarrassed by the oversight—obliged, enthusiastically endorsing Lemaître’s results: not before long would he and de Sitter fully acknowledge that Lemaître had hit on “a brilliant solution” (Eddington, as cited in Kragh, 2007, p.147) of the redshift enigma—as a manifestation of an \emph{expanding} universe. Eddington made sure that Lemaître’s work would be brought to the attention of the world at the next meeting of the Royal Society, and moreover, he sponsored an English translation.\par

Having studied more carefully studied Lemaître’s (1927) paper, Eddington re-analysed Einstein’s 1917 model. He showed that Einstein’s corresponding solution of his field equations was in fact unstable, ``like a pen balanced on its point” (Luminet, 2011, p.2914): a puny deviation from the delicately balanced matter distribution that Einstein had postulated would undo the model’s staticity; any arbitrarily small perturbation would cause the resulting universe to either contract or expand. This instability argument was the coffin nail for Einstein’s model: it exposed his static universe to require \emph{unphysical} fine-tuning. By the beginning of the next year, 1931, even Einstein converted to the dynamical universe (see Nussbaumer, 2014; McCoy, 2020, esp. sect. 2 for details). “By 1933, the theory of the expanding universe was accepted by a majority of astronomers and subjected to detailed reviews […]. It was also disseminated to the public through a number of popular works […]” (Kragh, 2007, p. 148). While the expanding universe had become a reality, it provoked new questions: what drives the expansion? How to improve our cosmological—from now on: typically dynamical—models? And how to test them?\footnote{ This question is of special importance since alternatives to models of an expanding universe were studied (e.g. Zwicky’s (1929) tired light hypothesis (expressly intended as an alternative explanation of the redshift-distance relation), Milne’s (1935) special-relativistic cosmological model, or the Steady State Cosmology of Bondi and Gold (1948))—even if only as \emph{minority} research programmes. O’Raifeartagh \& O’Keeffe (2020, p.222) rightly point out that “the redshifts of the nebulae are merely one manifestation of cosmic expansion; other manifestations exist, notably the frequency range of the cosmic microwave background [...].
} These would define the agenda of cosmology for the rest of the century until now (see e.g. Kragh, 2007, Ch. 4\&5; MacCallum, 2015; or the contributions in Kragh \& Longair, 2019). \par 

Who, then, discovered the expanding universe? In 2018, the International Astronomical Union voted to rename the redshift-distance relation, commonly known as “Hubble’s Law”. The resolution recommends “that from now on the expansion of the universe be referred to as the Hubble-Lemaitre law”. While the resolution passed with 78\% of voters in support, there was some pushback (Kragh, 2018b; Gr\o n, 2018; Elizalde, 2019, 2020, Ch.5; O'Raifeartaigh \& O'Keeffe, 2020). In part, it excoriated historical omissions and distortions (grievances that the above presentation, we hope, evades); in part, the criticism was based on different construals of the very notion of a discovery—without explicit attention to those construals themselves. The following subsections will fill this lacuna. To that end, we’ll first (\textbf{§4.2}) apply the philosophical accounts discussed in \textbf{§2} to the expanding universe case, and then turn to the change-driver model (\textbf{§4.3}). \par

\subsection{The view from extant accounts}

With the historical material in hand, let’s now apply the models of discovery (\textbf{§2.1}-\textbf{§2.4}) to the case of the expanding universe. Whom amongst our dramatis personæ of \textbf{§4.1} should we proclaim its discoverer? We’ll argue: none! Those models rule out the most plausible candidates, or they are simply inapplicable for reasons \emph{inherent} to those models.\par

First, what is the result on a \textbf{Kuhnian} analysis?\footnote{We’ll set aside a fundamental line of attack, criticising the applicability of Kuhn’s model of theory dynamics to cosmology (see Marx \& Bornmann, 2010; and O’Raifeartagh, 2013 for negative assessments, and Kragh \& Smith, 2003, pp.143; Kragh, 2007, pp.243 for more positive ones).} The single most important candidate ``what-that”-discovery of the expanding universe—a discovery where a phenomenon’s anticipated existence and conceptualisation predate its empirical confirmation—was Lemaître’s prediction of a linear redshift-distance relation. But since for Kuhn, discovery requires \emph{both} a theoretical and an empirical component, the applicability of Kuhn’s model falters: while Lemaître lacked the empirical data confirming his prediction, the person who provided the confirmation, Hubble (together with Humason), rejected the interpretation as a confirmation of the expanding universe (an epistemologically not unjustified attitude). Arguably the most plausible reaction, within Kuhn’s model, is to regard the episode as a discovery with joint (or collective) discoverers, none of whom can \emph{individually} be credited with the discovery.\footnote{Verbatim the same conclusion is reached for other “what-that”-discoveries concerning the expanding universe, such the predictions of the Alpher-Beta-Gamow theory.} The same conclusion is reached for the “that-what”-discovery of the expanding universe. While the researchers involved in the “anomalous redshift data” (chiefly Slipher, Hubble and Humason) lacked the theoretical framework for their findings, the person who provided that, Lemaître, lacked the empirical findings. \par

The Kuhnian perspective is disappointing in its reduction of the rich nexus of discoveries, reviewed in \textbf{§4.2}, to the above black and white conclusion; it belies the complexity of historical science.\par

Let’s move on to Schindler’s model. For \textbf{Schindler}, to discover means both to observe an object X or its direct effects and to conceptualise the essential properties needed to identify it.\par

A first stumbling stone to applying Schindler’s model to the expanding universe concerns quibbles regarding the requisite ontological category of the discovered entity. Cosmic expansion itself isn’t an object: it’s a \emph{global, large-scale property} of the universe.\par

On the one hand, although Wirtz, Lundmark, Hubble and Humason have plausible claims to first observing the redshift-distance relation, this falls short of having discovered the expanding universe. None held the right conceptualisation—the interpretation of the phenomenological redshift-distance correlations in terms of an expanding universe. On the other hand, Schindler’s model also disqualifies Friedmann and Lemaître. Even though they possessed the right conceptualisation, neither had observed the expansion (or its direct effects).\par

Perhaps one might reconsider Lemaître: after all, he understood the significance of Hubble’s results as evidence for his previously identified conceptualisation. But also this option runs afoul of another problem: Schindler’s insistence on essence-identification. We’d be hard-pressed to say how this condition would be satisfied here. Is cosmic expansion a natural kind? Does it have an essence? In which sense is expansion an essential property of the universe? As far as standard cosmology is concerned, any such invocation of essentialism seems a non-starter: universes \emph{needn’t} expand; in no sense is expansion necessary! Even as associated with a natural kind, expansion doesn’t fit the bill. Relativistic cosmological models are arguably more naturally classified according to the spacetime (or spatial) curvature, i.e. geometrically. But there is \emph{no} one-to-one link between a cosmological model’s curvature and whether it represents an expanding universe.\par
On \textbf{Hudson’s} account, discovering an object X requires being the first person to formulate an appropriate base description, and the latter’s material demonstration; here, the base description is supposed to ensure—as far as we know—that whenever it’s satisfied, X is present. Setting aside the limitation that Hudson’s account shares with Schindler’s (due to the account’s focus on “objects”), let’s parse out Hudson’s criteria for the expanding universe. A natural base description of the expansion of the universe would seem: 
\begin{quote}
 
The distance between any two objects that are not gravitationally bound (and interact only negligibly in all other ways) increases over time.
   
\end{quote}

Now what would have to be “materially demonstrated”? Recall that for him (paraphrasing Hudson, 2001, p.78), “to have discovered the expanding universe one need only possess enough conceptual resources to recognize its presence in \emph{a fairly reliable manner}.” The material demonstration should confer sufficient empirical-evidential credentials upon the base description: when it’s satisfied one should have sufficient reason to epistemically prefer X’s presence \emph{over alternative hypotheses}.\par
For two historical reasons this condition rules out the standard candidate discoverers: none of them had sufficient epistemic warrant for the base description’s adequacy over alternatives. \par
First, the methodological status of GR, and a fortiori cosmological models based on it, until the 1950s was flimsy (Eisenstaedt, 1989): the theory could hardly count as well-tested, even on astronomical scales (with the available tests probing only minor deviations from Newtonian theory). The empirical evidence in its favour was, at best, inconclusive. On cosmic scales: “[…] (i)ndeed, it could be said that cosmology did not truly constitute a test for the general theory of relativity in these years [1940-1955]” (O’Raifeartaigh, 2022, p.13, original emphasis). With the cosmic age problem one had a puissant reason to be leery of models of an expanding universe: until the 1950s, the main ones entailed an age of the universe below that of the Earth and the oldest known stars! Distrust in an expanding universe, insofar as it rested on GR’s validity at cosmic scales, was thus \emph{not irrational}!\par

Secondly, and relatedly, alternative explanations for the redshift-distance relation—the main empirical clue for an expanding universe until the 1950s or even 1960s—existed (see Kragh, 2017). Promptly following the publication of Hubble’s paper, Zwicky (1929) “advocated thinking of the redshift as the result of an interaction between photons and intervening matter rather than cosmic expansion […]” (Kirshner, 2003, p.11). In a gravitational analogue of the Compton effect, light loses its energy (“gets fatigued”) through scattering on its journey from the source to the receiver. Zwicky indeed calculated a linear redshift-distance relation—effectively reproducing Hubble’s phenomenological law—but construed now as an energy dissipation. Several variants of the “tired light”-hypothesis have been suggested (see e.g. Kragh, 2019, sect. 4.7). Admittedly, after WWII its popularity dwindled. Nonetheless, Kirshner (2003, p.11) concedes: “(t)he reality of cosmic expansion and the end of the ‘tired light’ has only recently [in the 90s and early 2000s!] been verified in a convincing way.”\footnote{For an assessment of an earlier verification---but still significantly after an expanding universe had already been elevated to the standard model in cosmology---see Norton (2023).}\par

Lastly—\textbf{McArthur’s} account, a liberalised version of Hudson’s, amalgamated with structural realism: McArthur allowed for discoveries even prior to material demonstrations—provided the discoverer harnesses \emph{already well-confirmed theoretical} relations in deriving a base description of something not yet observed.\par

Like Hudson’s model, the application of McArthur’s model scuppers on the historical standing of relativistic cosmology: GR wasn’t yet sufficiently well-tested. Accordingly, Friedmann’s or Lemaître’s derivations of the base description for cosmic expansion couldn’t rely on relations—GR’s structural content—that would qualify as well-confirmed. Not even the third of GR's so-called “classical tests” which are phenomenologically and explanatorily closest to the redshift-distance relation, gravitational redshift (i.e. energy loss of electromagnetic radiation in the presence of a gravitational field, manifesting itself in redshift) had been satisfactorily performed;\footnote{While Adams’ measurements of the star Sirius B, a white dwarf, in 1925 were viewed as (tentative) confirmatory evidence, a critical re-evaluation has deflated such a view (Hetherington, 1980; Holberg, 2010; see also Earman \& Glymour, 1980 for fallacies and subtleties in the derivation and interpretation of gravitational redshift).} a convincing test of gravitational redshift would have to wait until the Pound and Rebka experiment in 1960.\footnote{A similar conclusion is reached, with a curious twist, for a slightly different base description for an expanding universe (for which we take our cue from O’Raifeartaigh et al.‘s (2014) characterisation of Einstein’s cosmological bias in terms of his preference for an unchanging—rather than static—universe. This is compatible with stationarity. Equating “expansion” with “evolution” or “variation with time”, one may adopt the alternative base description: “The universe changes over the course of time, with its evolution being time-dependent; cosmic structures aren’t stationary.” In this case, a discovery of the expanding universe, on McArthur’s model, wouldn’t have been possible before the Big Bang Theory’s main rival, the Steady-State Theory, had become empirically discredited, again in the 1950s or 1960s (see Kragh, 1996, Ch. 6\&7).}\par

In conclusion, the three main models of discovery don’t apply to the case of the expanding universe. Per se, we don’t deem such a negative result—difficulties in identifying an individual discoverer—problematic. What \emph{does} seem problematic is rather the reason for those model’s inapplicability: their general \emph{philosophical} assumptions about what science can and does achieve. Unproblematically, by contrast, applying models of discovery may fail through \emph{historiographical complexities}: say, because many researchers contributed important elements or because the relevant theoretical and empirical developments took a long time, it may not be possible to single out one or several individuals.\par

\subsection{Applying the change-driver account}

The change-driver account yields a nuanced verdict on the discovery of the expanding universe. It's able to make specific claims, while doing justice to the complexity of the science.

\subsubsection{Empirical-physical discoveries of the expanding universe}

The empirical-physical discovery of the expansion of the universe begins with \textbf{Slipher} in the 1910s. His measurements suggested a curious empirical trend amongst faint spiral nebulae. In a list of velocities measurements (inferred from spectral displacements) published in 1914, “all but a few of the velocities were velocities of recession”, i.e. moving away from Earth/our galaxy. “In addition, the radial velocities of the spirals were generally much larger than the radial velocities of the stars or gaseous nebulae. The very size of the spectral shifts prompted some astronomers to query the Doppler, or velocity, interpretation of the shifts in the spectrum of the spirals” (Smith, 1979, p.136). Slipher, in other words, empirically-materially discovered the phenomenon that (i) the distances between us and other galaxies systematically tend to increase, and (ii) the velocities corresponding to this apparent “recession” are \emph{unusually} high (vis-$\grave{a}$-vis the proper motion of stars and nebulae).\footnote{Recall that Slipher’s pioneering velocity measurements were indispensable for Hubble’s (and Lemaître’s) own later work; Hubble’s ground-breaking achievement resulted from the combination of his own distance measurements and Slipher’s velocity measurements.
Interestingly, “(i)t is one of the great ironies of science that Hubble’s measurements of distance were later substantially revised due to significant systematic errors. Due to an error in the classification of Cepheid variables, Hubble’s cosmological distance ladder was later substantially revised by Walter Baade (1956) and Allan Sandage (1958)” (O’Raifeartaigh, 2013, p.8).
}\par

Slipher's work was careful and reasonably accurate, satisfying criterion (J) of the change-driver model and added something quite new to our understanding of the Universe (A). It also introduced an anomaly (P), namely the question of the cause of the seeming recession, which was quickly recognised. According to Eddington (1923, p.161), it constituted “one of the most perplexing problems of cosmogony”. Its absorption by extragalactic-astronomical background knowledge enabled new computational techniques (primarily for calculating solar motion), crucial for further developments. Soon attempts followed to correlate the radial velocities of those spiral galaxies with other of their observable parameters, in particular, their magnitudes and diameters (in turn correlated with distance).\par

The next milestone in the empirical-physical discovery of the expanding universe is the fruit of such endeavours: a positive correlation between the radial velocities and apparent magnitudes of the nebulae. Here it’s difficult to attribute this result definitively to an individual. “As early as 1918 Harlow Shapley […] had suggested that ‘the speed of spiral nebulae is dependent to some extent upon apparent brightness, indicating a relation of speed to distance, or possibly, to mass’, but he did not follow up this speculation” (Smith, 1979, p.142). Wirtz’s announcement that “he had established a well-defined observational relation between the radial velocities and apparent magnitudes of the spiral nebulae […] were not taken seriously by the principal students of the spiral nebulae” (ibid.); hence, not impacting the knowledge of the scientific community, the change-driver model disqualifies him, too, as a discoverer. \textbf{Lundmark}, too, found a relation between redshift and distance in his 1924 study. In his 1925 paper on “the motions and distances of the spiral nebulae” he even determined the relation to be dominantly linear. However, unlike other accounts of discovery, priority is not definitive for the change-driver model. Lundmark's determination was not justified, even by the standards of the day-he made errors and used only a single method to derive distances. Importantly for the change-driver model, Lundmark's work also failed to be taken up by his contemporaries.\par

The second major empirical-physical discovery relevant to our purposes is the full establishment of this linear relationship. We have already reviewed two anticipations of this discovery. Smith (1979, pp.147) writes, “Hubble, as he himself acknowledged, was working within a well-defined problem area, and before 1929 other astronomers had considered the existence of a redshift-distance relation. Hubble’s success was not to ‘discover’ a relation; rather, it was to \emph{convince} his colleagues that the relation was linear” (op.cit., p.133, emphasis in the original). However, contrary to Smith, we'd argue that Hubble's power to convince is highly relevant to the question of who (empirically-physically) discovered the expanding universe. On the change-driver model, Hubble's work counts as a discovery due to the superior quality of the data,  the impact Hubble’s (and Hubble and Humason’s) work had for the body of astronomical knowledge, and the quickly recognised relevance for the theoretical models of the universe. His use of multiple distance methods, his careful calculation, and his precise (for the time) data, justified his claims. And his evidence was taken up; it “ushered in a new area in cosmology by presenting a novel set of problems”, producing “a dialectic between theory and data in studies of the large-scale properties that was undreamt of only a few years before” (op.cit., p.157).\par

This leads us to the other motor of that dialectic, theory—and the associated \textbf{theoretical discoveries}, i.e. theory-internal, formal results related to the expanding universe.

\subsubsection{Theoretical discoveries concerning expanding universes}
The first milestone is best attributed to \textbf{Einstein and de Sitter} jointly, right at the dawn of relativistic cosmology in 1917. Einstein may be said to have formally discovered that his original (1915) field equations, sans cosmological constant, didn’t allow for a static solution—even if he \emph{rejected} that conclusion. To forestall it, he resorted to an ad-hoc modification of his original theory. Einstein’s (indirectly expressed) insight is complemented by de Sitter’s discovery of his “solution B”, the first non-static solution (even though its non-staticity was fully grasped only later). The object of Einstein’s and de Sitter’s twin formal discoveries were non-static cosmologies as theoretical possibilities within GR.  

That they are generic possibilities is the object of the second group of formal discoveries. Friedmann (1922) not only derived the equations governing cosmological dynamics from GR, but also explicitly showed that cosmological models \emph{typically} describe expanding universes. But given the delayed attention to his work, which—when it finally came—was mentioned in one breath with Lemaître’s independent re-discovery, this discovery is, on the change-driver model, most plausibly attributed to \emph{both} \textbf{Lemaître and Friedmann}, as independent co-discoverers. A second, impactful discovery in this second group is Eddington’s (1930) explicit demonstration of the instability of Einstein’s cosmological model. This exposed Einstein’s static universe to be fine-tuned or extraordinarily special (and thus the opposite of generic).\par

A third group of formal discoveries revolves around the linear redshift/distance relation as a \emph{consequence} of cosmological models. Here, simultaneously and independently of each other, \textbf{Weyl} (1923) and \textbf{Eddington} (1923) were the first to obtain the result in a solid mathematical way, even if it was limited to the de Sitter solution. \textbf{Lemaître} (1927) derived a more general result, for a generically expanding universe. A further refinement came a year later by Robertson: he derived the linear redshift/distance relation from the most general solution to Einstein’s field equations, subject to the constraints of homogeneity and isotropy.

\subsubsection{Synthetic discoveries of the expanding universe}
What synthetic discoveries does the change-driver model highlight—i.e. results that reveal an evidential relationship between empirical phenomena (already detected or still to be performed) and the notion of an expanding universe?  \par

We can discriminate between two distinct types, negative and positive synthetic discoveries. Members of the former pertain to evidential or explanatory deficiencies, members of the latter, to positive explanatory or evidential achievements.\par

\textbf{Negative synthetic discoveries} comprise two clusters. The first is composed of the waxing realisation that redshift data pose growing and intensifying anomalies for the existing cosmological models. It begins with Slipher’s redshift data, and continues with their solidification and the final emergence of a (linear) redshift-distance relation. The redshift data posed a perplexing challenge for static models of the universe: how to explain the trend towards redshift amongst galaxies? What sense to make of the high velocities (if one decided to translate, via an interpretation in terms of a Doppler effect, the redshift into velocities)? The redshifts weren’t predicted or easily explained within Einstein's static model. The explanatory resources of de Sitter’s model seemed more promising (based on minor positive synthetic discoveries, such as the various vague theoretical anticipations of a redshift-distance relation)—but the latter was patently unrealistic. Significant research efforts were invested into connecting de Sitter’s model with observed redshifts. Neither Einstein’s nor de Sitter’s model had the resources to explain the redshifts without additional suppositions. This, in the main, drove Lemaître to explore an expanding universe. It's inherently difficult to identify individual discoverers for the insight that anomalies aren’t naturally accommodated or explained within the existing cosmological models: different researchers must typically try out different ansätze—each of which turns out to be unsatisfactory. Plausibly, such negative synthetic discoveries are a \emph{collective} achievement: the explanatory inadequacy of the extant models gradually dawns on the scientific community.  \par

A second family of negative discoveries, often overlooked, concerns the evidential shortcomings of cosmological models representing an expanding universe, even after Hubble’s establishment of a linear redshift/distance relation, and after the widespread acceptance of the expanding universe. That the expanding universe models were evidentially underdetermined, with alternatives (\emph{later} to be refuted), plausibly counts as a significant insight that incentivised improved testing and the development of (unsuccessful) rivals (cf.\, for instance, Bschir, 2015).\footnote{Kragh (2019, p.120, also for a discussion of such rival theories) rightly stresses: “(t)hey belong to the history of the field as much as do the more successful theories that in a more direct way paved the way to the modern view of the universe.”} Again, several researchers can be given credit for such discoveries.  \par

On the side of \textbf{\emph{positive} synthetic discoveries}, one outshines all others: the insight that a linear redshift-distance relation receives a natural explanation—or is predicted—by an expanding universe. This insight can be attributed to \textbf{Lemaître}. He was the first to determine this empirical prediction of an expanding universe model, and identified the body of evidence that would, eventually, confirm that prediction. Of course, only Hubble’s result would corroborate it, and thereby actualise Lemaître’s otherwise merely \emph{potential} synthetic discovery. Only thanks to Eddington’s interposition did Lemaître’s discovery become a scientific discovery proper and advanced scientific knowledge within the physics community. Nonetheless, it was Lemaître’s interpretation of the redshift data that drove the acceptance—and further development of cosmology—of an expanding universe.\par

Hubble was well aware of that interpretation, but didn’t commit to it. To be sure, Hubble’s work, as an observational finding, had a huge influence on the acceptance of an expanding universe by virtue of Lemaître’s interpretation. On its own, it was an empirical-physical discovery, not a synthetic one: qua phenomenological law it didn’t link the notion of an expanding universe and observational phenomena. Hubble may, to some extent, be credited with a negative synthetic discovery: the \emph{underdetermination} of the expanding universe interpretation of the linear redshift-distance relation.\par

The importance of this negative discovery for the status of cosmology as an empirical science can’t be overstated. Still in 1963, “there (were) only 2.5 facts in cosmology” (Longair, 1993, p.160): the darkness of the night sky (peripheral to developments in cosmology at issue here), Hubble’s empirical redshift-distance law, and the half-fact, in the twilight since the first reliable source counts in the mid-1950s, “a matter of considerable controversy” (ibid.), concerning the universe’s evolution, i.e. the idea that ``the contents of the universe have probably changed as the universe grows older” (ibid.). Within a few years, cosmology multiplied the number of facts. Important tests were performed, confirming the model of an expanding universe (see e.g. Longair, 2019).

\section{Summary and conclusion}
The present paper undertook a comprehensive review of extant philosophical models/explications of scientific discovery. While Kuhn’s account is compromised by its entanglement with his wider, and problematic, views on science (especially incommensurability), we discern the principal weaknesses of the more philosophical accounts in their gratuitously strong metaphysical and epistemological commitments (especially to realism), and their disconnect from scientific change and normative-social aspects of discoveries.\par

As an alternative that avoids these defects, we proposed the change-driver model. It conceptualises discoveries as cognitive problems or problem-solutions that have significantly advanced science. In tandem with a more fine-grained taxonomy of forms of discovery---dependent on the kind of interaction between the discovery and existing knowledge---the change-driver model was shown to exhibit superior intensional and extensional adequacy over other models.\par

The case of the expanding universe further illustrated this, underscoring the model’s fertility and relevance for a recent science-political debate. Our model yields a nuanced verdict on whom to identify as a discoverer of what---a verdict that does greater justice to the complexity of science and its history than rivalling models. Chiming with other historical analyses, the change-driver model \emph{opposes} the IAU’s (2018) recommendation: while the empirical law in question—the correlation between redshift and distance of galaxies—ought to be attributed to Slipher, Hubble and (perhaps) Humason (rather than \emph{just} Hubble, or Hubble \emph{and Lemaître}), Lemaître (but not Hubble) deserves privileged credit for discovering its interpretation as evidence for the expanding universe. The wider theoretical understanding and epistemic backing of the claim that the universe expands is best viewed as a more collective accomplishment, involving several researchers. While both underpinned by, and in turn leading to numerous discoveries (see e.g.\ Longair, 2020), this extended (and in fact, on-going) process of consolidation doesn't qualify as a discovery itself; its epistemic function and anatomy is a different one.

\bibliography{sn-article.bib}%
\nocite{*}

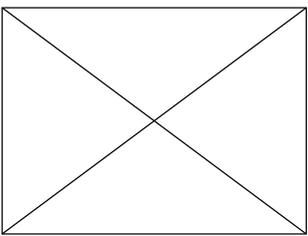
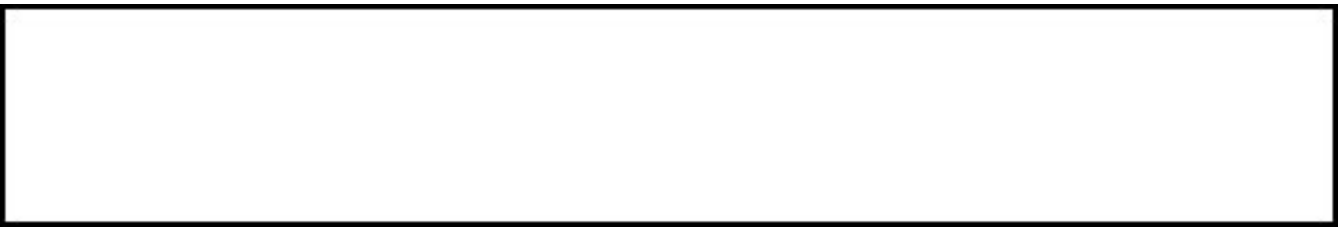
\end{document}